\documentclass[lettersize,journal]{IEEEtran}

\usepackage[utf8]{inputenc}
\usepackage{amsmath,amsfonts,amssymb}
\usepackage{amsthm}
\usepackage{bbding}
\usepackage{stmaryrd}
\usepackage{cite}
\usepackage{array}
\usepackage{textcomp}
\usepackage{stfloats}
\usepackage{tabularx}
\usepackage{threeparttable}
\usepackage{multirow}
\usepackage{booktabs}
\usepackage{url}
\usepackage{verbatim}
\usepackage[linesnumbered]{algorithm2e}
\usepackage{graphicx}
\usepackage{pifont}
\usepackage{wasysym}
\usepackage[colorlinks=true,linkcolor=blue,citecolor=blue,urlcolor=blue]{hyperref}
\usepackage{balance}
\usepackage{bm}
\usepackage[table]{xcolor}
\usepackage{colortbl}
\usepackage{subcaption}
\usepackage{enumitem}
\usepackage{multicol}
\usepackage{tikz}
\usepackage{xspace}

\usepackage{etoolbox}
\usepackage{marginnote}

\usepackage{rotating} % for sidewaystable

\newbool{showhighlight}  % 控制彩色高亮（revise/blue/red）
\setboolean{showhighlight}{true}   % true=彩色；false=不着色

\newbool{shownote}       % 控制边注（marginnote）
\setboolean{shownote}{true}     % true=显示；false=隐藏

\newbool{showtodo}       % 控制 TODO
\setboolean{showtodo}{true}        % true=显示；false=隐藏

\definecolor{R1color}{HTML}{ff8c00} % orange
\definecolor{R2color}{HTML}{762a83} % purple
\definecolor{R3color}{HTML}{01665e} % green

\newcommand{\one}[1]{%
  \ifbool{showhighlight}{\textcolor{R1color}{#1}}{#1}%
}
\newcommand{\two}[1]{%
  \ifbool{showhighlight}{\textcolor{R2color}{#1}}{#1}%
}
\newcommand{\three}[1]{%
  \ifbool{showhighlight}{\textcolor{R3color}{#1}}{#1}%
}

\let\oldmarginnote\marginnote
\renewcommand{\marginnote}[1]{%
  \ifbool{shownote}{\oldmarginnote{#1}}{}%
}

\newcommand{\revnote}[1]{%
  \ifbool{shownote}{\marginnote{\footnotesize #1}}{}%
}

\def\BibTeX{{\rm B\kern-.05em{\sc i\kern-.025em b}\kern-.08em
    T\kern-.1667em\lower.7ex\hbox{E}\kern-.125emX}}

\usepackage{wrapfig}
\usepackage{float}  % For [H] position specifier to prevent floating
\usepackage{listings}
\usepackage[most]{tcolorbox}

\newtcolorbox[auto counter]{mybox}[1][]{%
breakable,
enhanced,
sharp corners,
colback=white,
fonttitle=\bfseries,
enlarge bottom at break by=5mm,
enlarge top at break by=5mm,
overlay first={%
    \draw[black, line width=0.5mm](frame.south west)--(frame.south east);
    \node[anchor=north east] at (frame.south east) {continued on next page};
    },
overlay middle={%
    \draw[black, line width=0.5mm](frame.south west)--(frame.south east);
    \draw[black, line width=0.5mm](frame.north west)--(frame.north east);
    \node[anchor=north east] at (frame.south east) {continued on next page};
    \node[anchor=south west] at (frame.north west) {continued from next page};
    },
overlay last={%
    \draw[black, line width=0.5mm](frame.north west)--(frame.north east);
    \node[anchor=south west] at (frame.north west) {continued from next page};},
#1
}

  {\list{}{\leftmargin=0.15in\rightmargin=0.15in}\item[]}%
  {\endlist}

\usepackage{multicol}
\newtcolorbox{myquote}[1][]{
    colback=black!10,
    colframe=black!10,
    notitle,
    sharp corners,
    enhanced,
    breakable,
    left=2pt,
    right=2pt,
    top=2pt,
    bottom=2pt,
    ignore nobreak,
}

\definecolor{R1color}{HTML}{ff8c00} % orange
\definecolor{R2color}{HTML}{762a83} % purple
\definecolor{R3color}{HTML}{01665e} % green

\newtcolorbox{myquoteMeta}[1][]{
    colback=black!3,
    colframe=black!3,
    notitle,
    sharp corners,
    borderline west={1.5pt}{0pt}{blue},
    enhanced,
    breakable,
    left=2pt,
    right=2pt,
    top=0pt,
    bottom=0pt,
    ignore nobreak,
}

\newtcolorbox{myquoteR1}[1][]{
    colback=black!3,
    colframe=black!3,
    notitle,
    sharp corners,
    borderline west={1.5pt}{0pt}{R1color},
    enhanced,
    breakable,
    left=2pt,
    right=2pt,
    top=0pt,
    bottom=0pt,
    ignore nobreak,
}

\newtcolorbox{myquoteR2}[1][]{
    colback=black!3,
    colframe=black!3,
    notitle,
    sharp corners,
    borderline west={1.5pt}{0pt}{R2color},
    enhanced,
    breakable,
    left=2pt,
    right=2pt,
    top=0pt,
    bottom=0pt,
    ignore nobreak,
    #1
}

\newtcolorbox{myquoteR3}[1][]{
    colback=black!3,
    colframe=black!3,
    notitle,
    sharp corners,
    borderline west={1.5pt}{0pt}{R3color},
    enhanced,
    breakable,
    left=2pt,
    right=2pt,
    top=0pt,
    bottom=0pt,
    ignore nobreak,
}

\newtcolorbox{modificationR1}[2][]{
    colback=white,
    colframe=white,
    colupper=R1color,
    notitle,
    sharp corners,
    boxrule=0pt,
    enhanced,
    breakable,
    left=5pt,
    right=5pt,
    top=5pt,
    bottom=5pt,
    ignore nobreak,
    before upper={\underline{\textbf{Modifications in the manuscript:}} \textit{#2}\par\vspace{2mm}},
    #1
}

\newtcolorbox{modificationR2}[2][]{
    colback=white,
    colframe=white,
    colupper=R2color,
    notitle,
    sharp corners,
    boxrule=0pt,
    enhanced,
    breakable,
    left=5pt,
    right=5pt,
    top=5pt,
    bottom=5pt,
    ignore nobreak,
    before upper={\underline{\textbf{Modifications in the manuscript:}} \textit{#2}\par\vspace{2mm}},
    #1
}

\newtcolorbox{modificationMeta}[2][]{
    colback=white,
    colframe=white,
    colupper=blue,
    notitle,
    sharp corners,
    boxrule=0pt,
    enhanced,
    breakable,
    left=5pt,
    right=5pt,
    top=5pt,
    bottom=5pt,
    ignore nobreak,
    before upper={\underline{\textbf{Modifications in the manuscript:}} \textit{#2}\par\vspace{2mm}},
    #1
}

\usepackage{marginnote}
\usepackage{mparhack}

\newtheorem{theorem}{Theorem}

\theoremstyle{remark}
\newtheorem*{remark}{Remark}

\newcolumntype{C}[1]{>{\centering\arraybackslash}m{#1}}
\renewcommand{\arraystretch}{1.5}

\newcommand{\ignore}[1]{}

\begin{document}

\clearpage
\onecolumn
% % Use group to isolate cover_letter caption settings
% \begingroup
% \input{cover_letter}   % cover_letter 可以是多页
% \endgroup
% \clearpage

% Restore automatic numbering for manuscript body
\captionsetup[figure]{labelformat=default}
\captionsetup[table]{labelformat=default}

\twocolumn

% Reset figure and table counters to start from 1 in manuscript
\setcounter{figure}{0}
\setcounter{table}{0}

\title{DP2Guard: A Lightweight and Byzantine-Robust Privacy-Preserving Federated Learning Scheme for Industrial IoT}

\author{Baofu Han,
    Bing Li,
    Yining Qi,~\IEEEmembership{Member,~IEEE},
    Zhiquan Liu, ~\IEEEmembership{Senior Member,~IEEE},
    Raja Jurdak,~\IEEEmembership{Senior Member,~IEEE},
    Kaibin Huang, ~\IEEEmembership{Fellow,~IEEE},
    and Chau Yuen,~\IEEEmembership{Fellow,~IEEE}
\thanks{This work was supported in part by the Postgraduate Research \& Practice Innovation Program of Jiangsu Province under Grant KYCX23\_0320; in part by the Shenzhen Science Technology and Innovation Commission (SZSTI) under Grant JCYJ20170817115500476. (Corresponding author: Bing Li.)}
\thanks{Baofu Han is with the School of Cyber Science and Engineering, Southeast University, Nanjing 211102, China, and also with the School of Computing, National University of Singapore, Singapore (e-mail: hanbaofu@seu.edu.cn, baofu.han@comp.nus.edu.sg).}
\thanks{Bing Li is with the School of Microelectronics, Southeast University, Nanjing 211102, China (e-mail: bernie\_seu@seu.edu.cn).}
\thanks{Yining Qi is with the School of Computer Science and Technology, Huazhong University of Science and Technology, Wuhan 430074, China (e-mail: qiyining@hust.edu.cn).}
\thanks{Z. Liu is with the College of Cyber Security, Jinan University, Guangzhou 510632, China. (e-mail:  zqliu@jnu.edu.cn).}
\thanks{Raja Jurdak is with the School of Computer Science, Queensland University of Technology, Australia (e-mail: r.jurdak@qut.edu.au).}
\thanks{Kaibin Huang is with the Department of Electrical and Electronic Engineering, The University of Hong Kong, Hong Kong SAR, China (e-mail: huangkb@hku.hk).}
\thanks{Chau Yuen is with the School of Electrical and Electronic Engineering, Nanyang Technological University, Singapore (email: chau.yuen@ntu.edu.sg).}
}
\markboth{Journal of \LaTeX\ Class Files,~Vol.~18, No.~9, September~2020}%
{How to Use the IEEEtran \LaTeX \ Templates}

\maketitle

\begin{abstract}
Privacy-Preserving Federated Learning (PPFL) has emerged as a secure distributed Machine Learning (ML) paradigm that aggregates locally trained gradients without exposing raw data. To defend against model poisoning threats, several robustness-enhanced PPFL schemes have been proposed by integrating anomaly detection. Nevertheless, they still face two major challenges: (1) the reliance on heavyweight encryption techniques results in substantial communication and computation overhead; and (2) single-strategy defense mechanisms often fail to provide sufficient robustness against adaptive adversaries. To overcome these challenges, we propose DP2Guard, a lightweight PPFL framework that enhances both privacy and robustness. DP2Guard leverages a lightweight gradient masking mechanism to replace costly cryptographic operations while ensuring the privacy of local gradients. A hybrid defense strategy is proposed, which extracts gradient features using singular value decomposition and cosine similarity, and applies a clustering algorithm to effectively identify malicious gradients. Additionally, DP2Guard adopts a trust score–based adaptive aggregation scheme that adjusts client weights according to historical behavior, while blockchain records aggregated results and trust scores to ensure tamper-proof and auditable training. Extensive experiments conducted on two public datasets demonstrate that DP2Guard effectively defends against four advanced poisoning attacks while ensuring privacy with reduced communication and computation costs.
\end{abstract}

\begin{IEEEkeywords}
Blockchain, federated learning, poisoning attack, privacy preserving, hybrid-defense strategy.
\end{IEEEkeywords}

\section{Introduction}
\label{sec:introduction}
\IEEEPARstart {T}he Industrial Internet of Things (IIoT) connects various industrial devices through networks, enabling data collection, exchange, and analysis, and has played a significant role in advancing the digital transformation of industrial systems \cite{yang2021joint}, \cite{Cao10577218}, \cite{yang2025asynchronous}. Meanwhile, the integration of machine learning (ML) with IIoT enables industrial terminals to extract valuable insights from massive sensory data and make intelligent decisions in complex scenarios \cite{11368752}, \cite{11389150}, \cite{Arunan10128148}. However, the advancement of industrial intelligence heavily depends on large-scale data collection and sharing, which raises serious concerns regarding data privacy and security \cite{tang2025verifiable}, \cite{10121733, 11372457}. Thus, how to fully realise the potential of ML in IIoT systems while preserving data privacy has become a critical challenge. 

Federated Learning (FL) has emerged as a distributed Machine Learning (ML) paradigm that enables multiple organizations to collaboratively train a global model without sharing raw data, thereby preserving user privacy \cite{11345450}, \cite{wang2026shift}, \cite{you2022triple}, \cite{you2025framework}. However, recent studies indicate that FL still encounters challenges that adversaries can infer sensitive information from shared model updates \cite{You10399971}, \cite{11237210}, \cite{11192575}. To address the privacy leakage in FL, some privacy-preserving federated learning (PPFL) based on Differential Privacy (DP) \cite{wei2020federated}, \cite{Lu8843942}, Homomorphic Encryption (HE) \cite{11269813}, \cite{Han10819476}, \cite{yang10316678}, and Secure Multi-Party Computation (SMPC) \cite{Yang9889177}, \cite{Bi10574892}, \cite{XuGuowen}, \cite{Zhang10476635} have been proposed. DP achieves privacy guarantees by injecting calibrated noise into local model gradients, but this inevitably introduces a trade-off between privacy and model utility. SMPC allows participants to jointly compute global models without revealing their inputs, but the interaction processes of SMPC incur heavy communication burdens on clients. In contrast, HE enables secure computation on encrypted data and offers a compelling balance between privacy and accuracy.

\begin{figure}[htbp]
	\centering
	\includegraphics[width=1\linewidth]{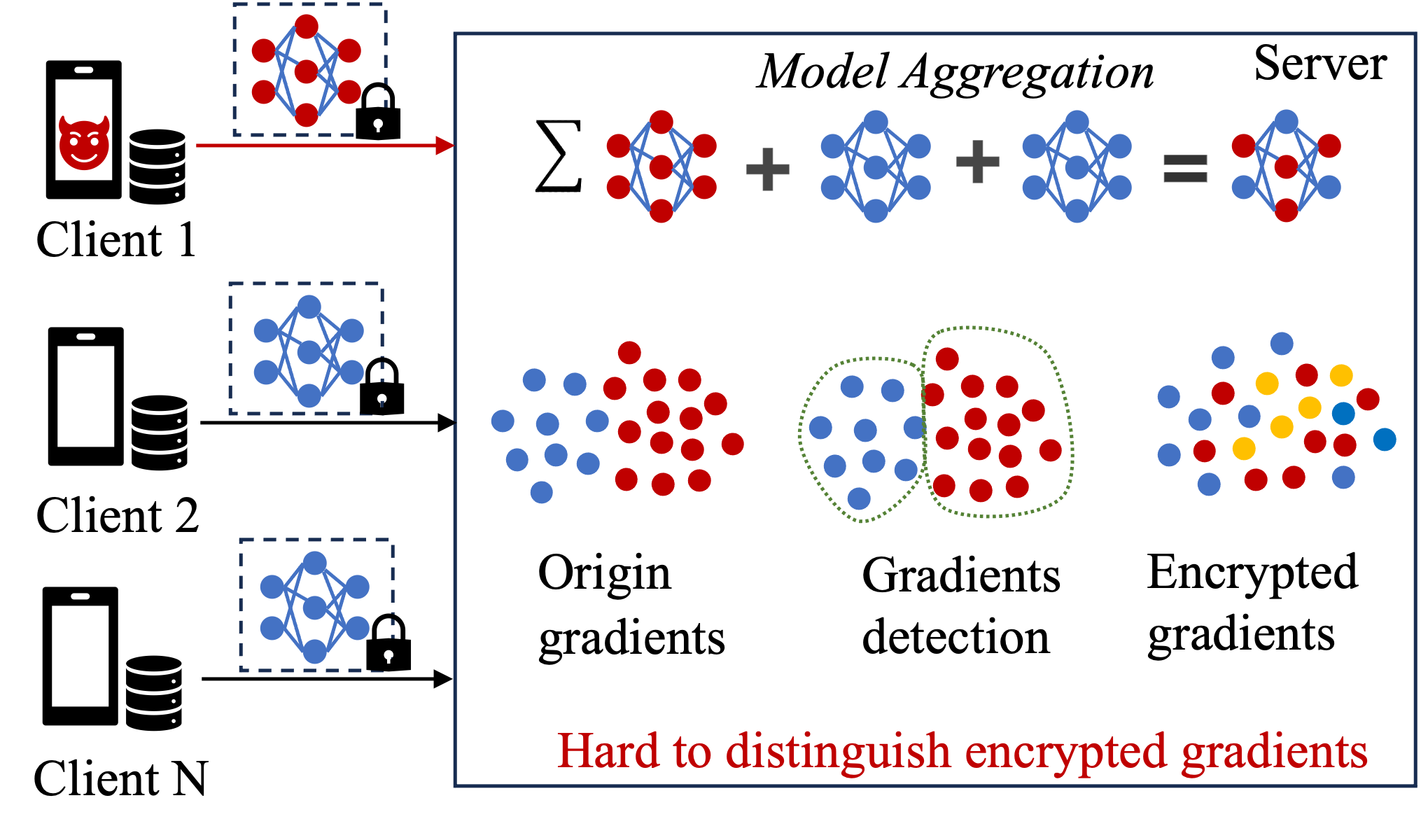}
	\caption{The example of model poisoning attacks in PPFL. Client 1 uploads a poisoned local mode gradient, guiding the global model towards a predefined direction and impacting overall model performance.}
	\label{Fig:PoisoningAttack}
\end{figure}

Another security threat in FL is model poisoning attacks (MPAs), in which malicious clients submit manipulated gradients to interfere with the training process and compromise the integrity of the global model \cite{Wang10049998}. As illustrated in Fig. \ref{Fig:PoisoningAttack}, such attacks become even more difficult to detect in PPFL due to the invisibility of individual gradients resulting from encryption or perturbation mechanisms. To address this issue, several defense strategies have been proposed, including cosine similarity-based \cite{Yuan10381871}, \cite{Fotohi10579800} and distance-based methods \cite{Zheng10930618}, \cite{dong2021flod}, \cite{Gehlhar10188630}, which identify abnormal updates by measuring their deviation from the majority. However, recent studies \cite{Kasyap10398506}, \cite{Wang10896961} have shown that under complex attacks-such as LIE \cite{baruch2019little} and Min-Max/Min-Sum \cite{shejwalkar2021manipulating}-attackers can craft malicious gradients that closely mimic benign behavior, thereby evading detection. Thus, single-strategy defenses are inadequate for addressing increasingly sophisticated poisoning threats, highlighting the need to develop more robust and hybrid defense mechanisms in PPFL.

To address the aforementioned challenges, we propose DP2Guard, a lightweight PPFL framework that enhances both privacy and robustness. DP2Guard adopts a gradient masking strategy to protect local model updates without relying on computationally expensive cryptographic techniques such as HE or MPC. It also incorporates a hybrid anomaly detection strategy to identify malicious updates. Blockchain is further integrated to secure the training process and provide auditability. The main contributions of this work are summarized as follows:

\begin{enumerate}
\item We propose a lightweight PPFL framework that utilizes efficient gradient masking to avoid the heavy overhead of cryptographic approaches. Furthermore, blockchain is integrated to enhance training security and auditability.

\item We develop a hybrid anomaly detection mechanism combining cosine similarity and spectral analysis, which more effectively detects malicious updates compared to conventional single-strategy defenses, and does so without accessing raw gradient data.

\item We introduce a trust score–based aggregation mechanism that dynamically adjusts client contributions based on their historical behavior to improve global model performance.

\item Extensive experiments on two public datasets demonstrate that DP2Guard provides strong resistance against four advanced poisoning attacks, while achieving privacy protection with reduced communication and computation costs.
\end{enumerate}

The remainder of this paper is organized as follows: Section \ref{sec:relatedwork} reviews related work in areas relevant to our study. Section \ref{sec:Preliminaries} introduces the preliminary knowledge. Section \ref{sec:SystemModel} illustrates the system model and the threat model. The proposed scheme is elaborated in Section  \ref{sec:OurScheme}. Section \ref{sec:Security} provides the security and privacy analysis. Section \ref{sec:Result} presents performance evaluation. Finally, Section \ref{sec:concludes} concludes this work.

\section{Related work}
\label{sec:relatedwork}

\begin{table*}[!t]
\caption{Existing Approaches Based on Malicious Client Detection in PPFL}
\label{tb:relatedwork}
\centering
\begin{tabularx}{\textwidth}{|C{2.1cm}|C{2.2cm}|C{2.2cm}|C{1.5cm}|C{1.5cm}|C{1.5cm}|C{1.5cm}|C{2.15cm}|}
\hline
\textbf{Solution} & \textbf{Privacy}& \textbf{Detection approach} & \textbf{Adaptive Aggregation} & \textbf{Test dataset} & \textbf{Blockchain} & \textbf{Byzantine Robustness} & \textbf{Distribution setting} \\
\hline
FLTrust \cite{cao2020fltrust} &\Circle & Cosine similarity& \Circle & \Circle & \Circle & \LEFTcircle& IID and Non-IID \\
Multi-Krum \cite{Blanchard} &\Circle& Krum & \Circle & \CIRCLE & \Circle & \Circle & IID \\
DnC \cite{shejwalkar2021manipulating} &\Circle& SVD & \LEFTcircle & \CIRCLE & \Circle & \LEFTcircle & IID \\
RFLP \cite{Li10233385} &HE& Krum & \LEFTcircle & \CIRCLE &  \Circle & \LEFTcircle & IID and Non-IID \\
Biscotti \cite{Sha9292450} &DP& Multi-Krum & \LEFTcircle & \CIRCLE & \CIRCLE & \LEFTcircle & IID \\
RFLPA \cite{mai2024rflpa} &HE& Cosine similarity & \LEFTcircle & \CIRCLE & \Circle  & \LEFTcircle & IID and Non-IID \\
ShieldFL \cite{MaZhuoran} & Double-trapdoor HE & Cosine similarity & \LEFTcircle & \CIRCLE & \CIRCLE & \LEFTcircle & IID and Non-IID \\
FLOD \cite{dong2021flod} & SMPC & Hamming distance & \Circle & \Circle & \Circle & \Circle & IID \\
PEFL \cite{Liu9524709} & HE & Statistical correlation-based & \Circle & \CIRCLE  & \Circle & \Circle & IID \\
VerifyNet \cite{XuGuowen} & VSS & No & \Circle & \Circle  & \Circle & \Circle & IID \\
DPFLA \cite{feng2024dpfla} &Removable mask& SVD & \Circle & \CIRCLE & \Circle & \LEFTcircle & IID \\
\textbf{DP2Guard} &Mask&Hybrid strategy& \CIRCLE & \CIRCLE & \CIRCLE & \CIRCLE & IID and Non-IID \\
\hline
\end{tabularx}

\vspace{1mm}
\small
\noindent * The symbol \CIRCLE\ indicates YES, \Circle\ indicates NO, \LEFTcircle\ indicates Partially YES. In the column of Byzantine Robustness and Adaptive Aggregation, the half circle \LEFTcircle\ indicates moderate support or defense. A full circle \CIRCLE\ indicates strong support/robustness.

\end{table*}

PPFL employs advanced techniques such as DP, HE, and SMPC to protect the confidentiality of model updates. Jiang et al. \cite{Jiang9628062} proposed a DP-based PPFL framework with adaptive gradient compression to enhance privacy while reducing communication overhead. Gu et al. \cite{Gu10324410} introduced FL2DP, which leverages exponential-based noise and gradient shuffling to protect both gradient and identity privacy. Li et al. \cite{Li10637763} designed an adaptive noise injection strategy to balance privacy protection with model accuracy. However, DP-based methods inevitably introduce utility loss, making it difficult to optimize both privacy and performance. To improve efficiency, SMPC-based PPFL approaches have been explored. 
Bonawitz et al. \cite{bonawitz2017practical} introduced a secure aggregation scheme using double masking with Shamir’s Secret Sharing (SSS) to protect client data privacy during aggregation. 
Xu et al. \cite{XuGuowen} proposed VerifyNet, which integrates homomorphic hash-based verifiability with secret sharing to protect gradient privacy and handle user dropout.
Nevertheless, such methods still suffer from increasing communication overhead as the number of clients grows. HE-based approaches offer stronger privacy guarantees by enabling computations on encrypted data. 
Li et al. \cite{Li9795898} utilized a CKKS cryptosystem to protect model parameters during training. However, HE-based approaches incur substantial ciphertext expansion and computational overhead.

To address the poisoning attacks, numerous Byzantine-robust algorithms have been proposed. For instance, Cao et al. \cite{cao2020fltrust} introduced FLTrust, which calculates the cosine similarity between local and root gradients. However, this approach relies heavily on a root dataset. DnC \cite{shejwalkar2021manipulating} adopts random projection for dimensionality reduction, followed by spectral analysis using SVD to identify and filter out malicious clients. However, DnC require a preset number of malicious clients, leading to limitations in practical applications. To improve privacy, Shayan et al. \cite{Sha9292450} proposed Biscotti, a robust PPFL scheme that combines the Multi-Krum \cite{Blanchard} strategy with differential privacy, detecting anomalous updates based on Euclidean distances. However, these strategies require access to plaintext local models. Li et al. \cite{Li10233385} proposed a robust PPFL based on Krum to defend against model poisoning attacks under encrypted gradient aggregation. Mai et al. \cite{mai2024rflpa}  introduced RFLPA, which integrates cosine similarity detection with verifiable secret sharing to ensure secure aggregation. 
Ma et al. \cite{MaZhuoran} proposed ShieldFL, utilizing double-trapdoor HE and secure cosine similarity to identify malicious clients. 
Liu et al. \cite{Liu9524709} introduced PEFL based on partially homomorphic encryption, which computes the coordinate-wise median as a benchmark and excludes updates with low Pearson correlation to the benchmark.
Feng et al. \cite{feng2024dpfla} introduced DPFLA, combining removable masks with SVD-based techniques for malicious gradient filtering, yet it assumes a non-colluding single server and remains vulnerable to collusion attacks. Overall, most existing robust and privacy-preserving FL schemes depend on a single defense mechanism, and as adversarial tactics become more adaptive and sophisticated, their robustness guarantees remain limited.

In Table \ref{tb:relatedwork}, we provide a comprehensive summary of the state-of-the-art robust PPFL approaches and compare them with our proposed scheme. The comparison highlights the advantages of our design in terms of privacy preservation, Byzantine robustness, and computational efficiency.

\section{Preliminaries}
\label{sec:Preliminaries}
In this section, we provide background on federated learning and briefly review state-of-the-art poisoning attacks.

\subsection{Federated Learning}

Assume there are $N$ clients, each of them owns a private dataset $D_i$, where $i \in[1, N]$. The objective of FL is to learn a global model $\omega$ that minimizes the overall global loss across all clients \cite{McMahan}:
\begin{equation}
    \min _\omega \sum_{i=1}^N \frac{\left|D_i\right|}{\sum_{j=1}^N\left|D_j\right|} \cdot \mathcal{L}_i(\omega),
\end{equation}
where $\mathcal{L}_i(\omega)=\frac{1}{\left|D_i\right|} \sum_{x \in D_i} \ell(\omega, x)$, and $\ell(\omega, x)$ denotes the loss function evaluated on sample $x$ using model $\omega$.

At the $t$-th iteration, the server distributes the current global model $\omega_t$ to all clients. Each client $i$ performs local training by minimizing its local loss $\mathcal{L}_i\left(\omega_t\right)$ using a local optimizer (e.g., stochastic gradient descent), and computes the gradient:
\begin{equation}
    g_i^{(t)}=\nabla_\omega \mathcal{L}_i\left(\omega_t\right) .
\end{equation}

The client then updates its local model as $\omega_i^t=\omega_t-\eta g_i^{(t)}$, where $\eta$ is the learning rate. The resulting model update $\Delta_i^{(t)}=\omega_i^t-\omega_t$ is sent back to the server. The server aggregates the received updates using FedAvg algorithm \cite{McMahan}:
\begin{equation}
    \omega_{t+1}=\omega_t+\frac{1}{\left|S_t\right|} \sum_{i \in S_t} \Delta_i^{(t)}
\end{equation}

This process is repeated until the global model converges or satisfies a predefined stopping criterion.
\subsection{Poisoning Attacks on FL}
Due to the decentralized nature of FL and the inability of the server to access raw client data or verify local updates, adversaries may interfere with the model aggregation process by injecting mislabeled data or constructing malicious model updates. These attacks are typically classified as non-adaptive or adaptive, depending on whether the perturbation is adjusted during training.

1) Non-Adaptive Attack: Non-adaptive attacks generate malicious updates using fixed strategies, regardless of their actual effect on the model. A common method is the label-flipping attack \cite{fung2020limitations}, where the attacker replaces a sample's truth label $y$ with an adversarial label $\hat{y}$, computed as:
\begin{equation}
    \hat{y}=y+l_{\text {flip }} \bmod L,
\end{equation}
where $L$ is the total number of classes and $l_{\text {flip }}$ is a fixed offset. The poisoned gradients are then uploaded for aggregation.

2) Adaptive Attack: Adaptive attacks dynamically adjust the perturbation direction or magnitude to improve both the effectiveness and stealth of the attack. Three representative adaptive attack strategies \cite{fang2020local}, \cite{shejwalkar2021manipulating}  are outlined below.

\begin{itemize}
    \item Fang attack \cite{fang2020local}: In this attack, the attacker simulates the aggregation rule using known benign gradients and inserts perturbations in the opposite direction:
    \begin{equation}
        \hat{g}_m=\frac{\sum_{i=1}^k g_i}{k}-\lambda * \operatorname{Sign}\left(\frac{\sum_{i=1}^k g_i}{k}\right),
    \end{equation}
    where $\lambda$ is the perturbation factor, and $g_i$ represents benign gradients. The attacker iteratively tunes $\lambda$ to ensure that the crafted gradients are accepted by the target aggregation rule.

    \item Min-Max \cite{shejwalkar2021manipulating}: The Min-Max attack aims to maximize the separation between the malicious gradient and benign gradients, while keeping the distance within the natural variation among benign clients. The formulation is:
    \begin{equation}
        \begin{aligned}
            & \arg \max _\gamma \max _{i \in[1, n]}\left\|\hat{g}_m-g_i\right\|_2 \leq \max _{i, j \in[1, n]}\left\|g_i-g_j\right\|_2, \\
            & \hat{g}_m=\frac{\sum_{i=1}^k g_i}{k}+\gamma \nabla_p,
        \end{aligned}
    \end{equation}

    where $n$ is the total number of clients, $\nabla_p$ is a predefined perturbation direction, and $\gamma$ is a scaling factor optimized by the adversary.

    \item Min-Sum \cite{shejwalkar2021manipulating}: In the Min-Sum attack, the goal is to minimize the cumulative distance between the malicious update and all benign gradients, thereby increasing its similarity to the majority and reducing its detectability. The optimization objective is:
    \begin{equation}
        \hat{g}_m=\arg \min _{\hat{g}} \sum_{i=1}^n\left\|\hat{g}-g_i\right\|_2+\gamma \cdot \nabla_p
    \end{equation}
    where $\gamma \cdot \nabla_p$ introduces controlled perturbation in the direction determined by the attacker.
\end{itemize}

\section{System and Threat Model}
\label{sec:SystemModel} 

In this section, we will describe the system model, threat model, and design goals, respectively.

\subsection{System Model}

\begin{figure}[htbp]
	\centering
	\includegraphics[width=1\linewidth]{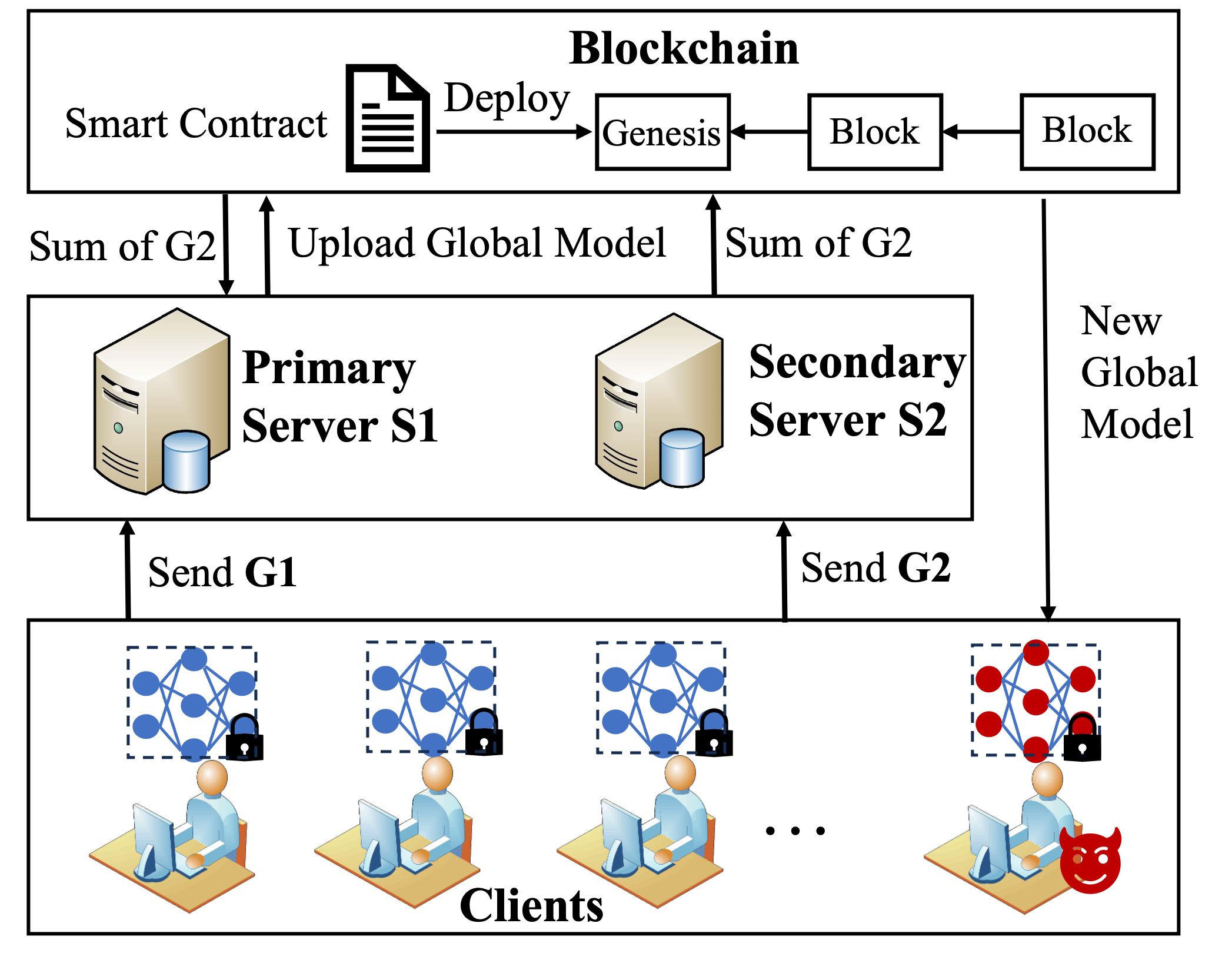}
	\caption{System architecture of the DP2Guard.}
	\label{Fig:BFL}
\end{figure}
As illustrated in Fig. \ref{Fig:BFL}, DP2Guard is a privacy-preserving and Byzantine-robust FL framework designed for IIoT environments. It adopts a dual-server structure integrated with a blockchain to ensure secure, reliable, and auditable aggregation. In each training round, an edge device $i$ performs local training and splits its computed gradient $G_i$ into two masked parts, denoted as $G_i^1$ and $G_i^2$. The Secondary Server (S2) performs hybrid anomaly detection to identify malicious updates and aggregates the $G_i^2$ shares. This aggregated result is then written to the blockchain to prevent tampering. The Primary Server (S1) retrieves the aggregated gradient from the blockchain and combines it with the sum of $G_i^1$ to reconstruct the global update. The updated model is subsequently  distributed to all devices via the blockchain, enabling efficient and trustworthy collaboration in IIoT network.

\subsection{Threat Model}
We consider two types of clients: (1) malicious clients; and (2) curious-but-honest clients. Malicious clients deliberately compromise the training process by submitting poisoned gradients aimed at degrading the global model’s accuracy. We assume that the maximum proportion of malicious clients does not exceed half of the total participating devices. In contrast, curious-but-honest clients follow the protocol but may attempt to infer private information from shared local updates, and potentially collaborate with others to improve inference success. 
The aggregation process is jointly managed by two servers, S1 and S2, which are assumed to be non-colluding and curious-but-honest—that is, they faithfully execute the protocol but may analyze collected data to recover individual client updates.
For the blockchain, we follow a standard Byzantine fault tolerance (BFT) model commonly employed in permissioned blockchain systems \cite{reijsbergen2022securing}, \cite{mollah2020blockchain}, where at least $\zeta > 2/3$ of trustee nodes are trustworthy and consistently return correct computation results. For simplicity, we assume all trustees have uniform computational power.

\subsection{Design Goal}
Given the aforementioned security and privacy threats in FL, DP2Guard is designed to meet four key goals: 

\begin{enumerate}
    \item \textit{Accuracy:} The scheme should preserve the accuracy of the global model when no attacks are present. Specifically, the integration of privacy-preserving and anomaly detection mechanisms should not negatively impact the model’s ability to learn from legitimate client updates.
    \item \textit{Robustness:} The scheme should exhibit strong resilience against various types of poisoning attacks, including more complex and adaptive ones. It must be capable of detecting or mitigating malicious updates effectively, while preserving the privacy of benign clients.
    \item \textit{Privacy:} The scheme must guarantee the privacy of each client’s local updates, ensuring that no adversary (e.g., the servers or other clients) can infer sensitive information from the shared local updates. Only the originating client should have access to the underlying private data.
    \item \textit{Efficiency:} Given the limited computational and communication resources of edge devices, the scheme should be designed with minimal overhead under privacy-preserving requirements, making it suitable for deployment on resource-limited clients. 
\end{enumerate}

\section{DP2Guard Algorithms and Architecture}
\label{sec:OurScheme}

In this section, we first introduce the overview of DP2Guard scheme. Then, we elaborate on the detailed design of DP2Guard. The main notations are shown in TABLE \ref{tab:notations}.

\begin{table}[ht]
\caption{Notation and Description}
\label{tab:notations}
\renewcommand{\arraystretch}{1.15}
\setlength{\tabcolsep}{3pt}
\centering
\begin{tabular}{>{\centering\arraybackslash}m{0.13\linewidth} p{0.32\linewidth} >{\centering\arraybackslash}m{0.13\linewidth} p{0.32\linewidth}}
\toprule
\textbf{Notation} & \textbf{Description} & \textbf{Notation} & \textbf{Description} \\
\midrule
$E_i$ & Client $i$ & $D_i$ & Local dataset of $E_i$ \\
$\mathbf{g}_i$ & Local gradient & $\tilde{\mathbf{g}}_i^{(1)}$/$\tilde{\mathbf{g}}_i^{(2)}$ & Masked gradient \\
$\omega_{t}$ & $t$-th round global model & $\omega_t^i$ & $t$-th round local model  \\
$\mathbf{r}_i$ & Random mask value & $\hat{\mathbf{g}}_i^{(1)}$/$\hat{\mathbf{g}}_i^{(2)}$& Mean-centered gradient \\
$\eta$ & Learning rate & $\mathbf{G}$ & Matrix of centered gradients \\
$s_i$ &  Spectral deviation score & $ \mathbf{f}_i$ & Feature vector of client $E_i$ \\
$c_i$ &  Median cosine similarity score & $\operatorname{Trust}_i^{(t)}$ & Trust score\\
$\tau_i^{(t)}$ & Aggregation weight& $\tilde{\mathbf{g}}_{\mathrm{agg}}^{(1)}$/$\tilde{\mathbf{g}}_{\mathrm{agg}}^{(2)}$ &Aggregated masked gradients on $\mathcal{S}_1$ and $\mathcal{S}_2$ \\
\bottomrule
\end{tabular}
\end{table}

\subsection {Overview}
\label{sec:overview}
\begin{figure*}[htbp]
	\centering
        \includegraphics[width=0.9\linewidth, height=13cm]{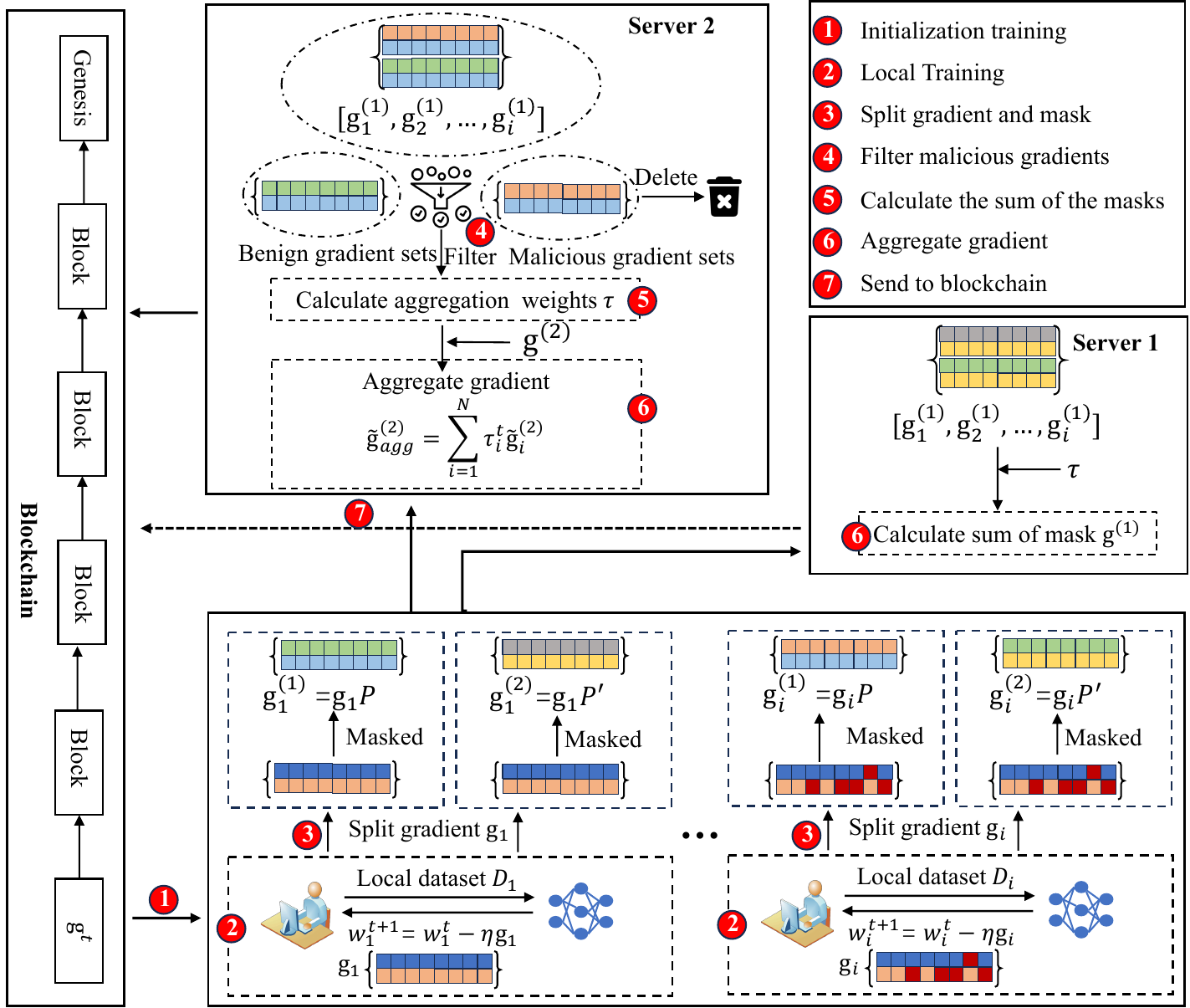}
	\caption{A brief workflow of the DP2Guard.}
	\label{Fig:flow}
\end{figure*}

To mitigate both privacy leakage and gradient poisoning in FL under IIoT scenarios, we propose DP2Guard, a dual-server defense framework that ensures both privacy preservation and Byzantine robustness. As illustrated in Fig. \ref{Fig:flow}, DP2Guard integrates gradient splitting and masking with hybrid anomaly detection to achieve secure model aggregation.

During the local training phase, each edge device $i$ computes a gradient vector $\mathbf{g}_i$ based on its private data. The client then splits $g_i$ into two additive shares, $\mathbf{g}_i^{(1)}$ and $\mathbf{g}_i^{(2)}$, such that $\mathbf{g}_i = \mathbf{g}_i^{(1)} + \mathbf{g}_i^{(2)}$. Each share is independently masked with a random vector $r_i$, forming $\tilde{\mathbf{g}}_i^{(1)} = \mathbf{g}_i^{(1)} + r_i$ and $\tilde{\mathbf{g}}_i^{(2)} = \mathbf{g}_i^{(2)} - r_i$, which are transmitted to S1 and S2, respectively.

Upon receiving the masked shares from all clients, S1 performs mean-centering on the collected $\tilde{\mathbf{g}}_i^{(1)}$ vectors and forwards the processed results to S2. Leveraging both the centered gradients and the locally held $\tilde{\mathbf{g}}_i^{(2)}$, S2 executes a hybrid anomaly detection mechanism to identify malicious behaviors, calculates trust scores for each client, and conducts trust-weighted aggregation. The resulting global update and trust weights are recorded on the blockchain to ensure transparency and tamper resistance. Finally, S1 downloads the aggregated gradients and trust weights from the blockchain and completes the final aggregation using its own masked shares.

\subsection{Design of DP2Guard}
\label{sec:design}

DP2Guard consists of four main phases: task initialization, local model training, privacy-preserving hybrid defense, and adaptive aggregation. Below, we provide a detailed description of each phase.

\textbf{1) Task Initialization:} In this phase, the task publisher  initializes the FL process by defining an initial global model, denoted as $\mathbf{w}^{(0)}$, which is then published to the blockchain to guarantee verifiability, integrity, and consistency across all participating devices. A subset of eligible mobile edge devices is then selected to participate in the training. Each selected device downloads the initial global model parameters $\mathbf{w}^{(0)}$ from the blockchain, which serve as the starting point for local training. 

\textbf{2) Local Model Training:} In the $t$-th communication round, each edge device $E_i$ updates its local model by training on its private dataset using the global model parameters $\mathbf{w}^{(t-1)}$ retrieved from the blockchain. After completing local training, the device computes its local gradient update $\mathbf{g}_i^{(t)}$. To protect the privacy of this gradient, DP2Guard designs a gradient splitting and masking strategy. Specifically, edge device $E_i$ first splits its gradient $\mathbf{g}_i$ into two additive shares $\mathbf{g}_i^{(1)}$ and $\mathbf{g}_i^{(2)}$, such that $\mathbf{g}_i=\mathbf{g}_i^{(1)}+\mathbf{g}_i^{(2)}$. 
To obfuscate the true gradient values, each device independently generates a fresh random mask vector $\mathbf{r}_i \in \mathbb{R}^{m \times n}$ at each communication round using a pseudo-random number generator (PRNG), where each element is independently and uniformly sampled. 
The mask generation is performed entirely locally on the device, without requiring any key agreement or coordination with other clients or servers.
Using this mask vector, the device creates two masked gradient shares:
\begin{equation}
    \tilde{\mathbf{g}}_i^{(1)}=\mathbf{g}_i^{(1)}+\mathbf{r}_i, \quad \tilde{\mathbf{g}}_i^{(2)}=\mathbf{g}_i^{(2)}-\mathbf{r}_i.
\end{equation}

These masked gradients are transmitted separately to Servers S1 and S2. This mechanism effectively safeguards against direct gradient leakage, no single server can infer the original gradient or any intermediate component without collaboration. The detailed procedure for this gradient splitting and masking process is presented in Algorithm \ref{alg:Training}.

\begin{remark} 
Since each client generates its mask vector independently and locally, no inter client communication or key negotiation is required, resulting in $O(N)$ communication complexity, which is significantly more efficient than secure aggregation protocols (e.g., SecAgg \cite{bonawitz2017practical}) that require pairwise key agreement with $O\left(N^2\right)$ overhead. 
Furthermore, a fresh mask vector is independently generated at each round, ensuring statistical independence across rounds and preventing attackers from recovering true gradients through multi-round differential analysis. 
Specifically, for any two rounds $t_1$ and $t_2$, the difference $\tilde{\mathbf{g}}_i^{\left(1, t_1\right)}-\tilde{\mathbf{g}}_i^{\left(1, t_2\right)}=\left(\mathbf{g}_i^{\left(t_1\right)}-\mathbf{g}_i^{\left(t_2\right)}\right)+\left(\mathbf{r}_i^{\left(t_1\right)}-\mathbf{r}_i^{\left(t_2\right)}\right)$ remains masked by the difference of two independent random vectors, which cannot be eliminated.
\end{remark}

\RestyleAlgo{ruled}

\SetKwComment{Comment}{/* }{ */}
\SetKwInOut{Input}{Input}\SetKwInOut{Output}{Output}

\begin{algorithm}
\caption{Local Model Training}\label{alg:Training}
\Input{Global model $\boldsymbol{w}^{(t-1)}$, Dataset $D_i$, Local gradient matrix  $\mathbf{g}_i^{(t)} \in \mathbb{R}^{m \times n}$}
\Output{Masked gradients $\tilde{\mathbf{g}}_i^{(1)}, \tilde{\mathbf{g}}_i^{(2)}$}
\textbf{Initialization:} $\boldsymbol{w}_i^{r}=\boldsymbol{w}^{r-1}$\;
\Comment{Training Model and Updating Gradient}
 \For{each edge device $i$}{
 Compute local gradient:
 $\mathbf{g}_i^{(t)} \leftarrow \nabla \ell\left(\boldsymbol{w}^{(t-1)} ; D_i\right)$\;
 Update local model: $\boldsymbol{w}_i^{(t)} \leftarrow \boldsymbol{w}^{(t-1)}-\eta \cdot \mathbf{g}_i^{(t)}$\;

\Comment{Gradient Splitting and Masking}
    Generate a random matrix \( \mathbf{r}_i \in \mathbb{R}^{m \times n} \) \;
    Generate two gradient shares $\mathbf{g}_i^{(1)} \in \mathbb{R}^{m \times n}$ and $\mathbf{g}_i^{(2)} \in \mathbb{R}^{m \times n}$\;
    Compute masked share $\tilde{\mathbf{g}}_i^{(1)} \leftarrow \mathbf{g}_i^{(1)}+\mathbf{r}_i$\;
    Compute masked share $\tilde{\mathbf{g}}_i^{(2)} \leftarrow \mathbf{g}_i^{(2)} - \mathbf{r}_i$\;
\Comment{Sending the Masked Gradient to S1 and S2}
    Send  $\tilde{\mathbf{g}}_i^{(1)}$ to server S1 \;
    Send  $\tilde{\mathbf{g}}_i^{(2)}$ to server S2 \;
}

\end{algorithm}

\textbf{3) Privacy-Preserving Hybrid Defense:} During each training round, an honest edge device $E_i$ randomly samples a mini-batch $d_i$ from its local dataset $D_i$, and computes a benign gradient $g_i$ accordingly. In contrast, malicious clients may submit adversarial gradients deliberately crafted to impair the performance of the global model. This behavior can be formally described as: 
\begin{equation}
    \mathbf{g}_i^t= \begin{cases}\nabla \mathcal{L}\left(\boldsymbol{w}_i^{t-1}, d_i\right), & \text { if device } i \text { is honest} \\ \mathcal{A}_{\text {attack }}\left(t, \mathcal{G}_{\text {honest }}, \Theta\right), & \text { if device } i \text { is malicious }\end{cases}
\end{equation}
where $\mathcal{A}{\text{attack}}(\cdot)$ denotes an attack-generation function that may utilize the current training round $t$, the set of gradients from honest clients $\mathcal{G}{\text{honest}}$, and auxiliary parameters $\Theta$.

While traditional defense schemes (e.g., Euclidean distance \cite{cao2020fltrust}, cosine similarity \cite{Blanchard}) are effective in simple scenarios, they may struggle under adaptive poisoning attacks. Furthermore, such methods usually require access to raw gradient data, introducing significant privacy risks. To mitigate this, DP2Guard introduces a privacy-preserving hybrid defense mechanism that enables malicious gradient detection without compromising data privacy. The entire process is presented in Algorithm \ref{alg:byzantine_detection}, which consists of the following key steps:

1) Mean-Centering Gradient: Both servers independently perform mean-centering on the masked gradient shares they receive. Specifically, Server $S_1$ computes the centered vectors $\hat{\mathbf{g}}_i^{(1)}$ by subtracting the mean of all received masked shares $\tilde{\mathbf{g}}_i^{(1)}$. Likewise, $S_2$ processes $\tilde{\mathbf{g}}_i^{(2)}$:
\begin{equation}
    \hat{\mathbf{g}}_i^{(1)} = \tilde{\mathbf{g}}_i^{(1)} - \overline{\tilde{\mathbf{g}}}^{(1)}, \quad \text{where} \quad \overline{\tilde{\mathbf{g}}}^{(1)} = \frac{1}{N} \sum_{j=1}^N \tilde{\mathbf{g}}_j^{(1)}
\end{equation}
\begin{equation}
    \hat{\mathbf{g}}_i^{(2)} = \tilde{\mathbf{g}}_i^{(2)} - \overline{\tilde{\mathbf{g}}}^{(2)}, \quad \text{where} \quad \overline{\tilde{\mathbf{g}}}^{(2)} = \frac{1}{N} \sum_{j=1}^N \tilde{\mathbf{g}}_j^{(2)}
\end{equation}

Server $S_1$ then forwards the centered results $\hat{\mathbf{g}}_i^{(1)}$ to $S_2$, which conducts hybrid detection without gaining access to any individual client’s raw updates.

2) Gradient Reconstruction: Upon receiving $\hat{\mathbf{g}}_i^{(1)}$ from $S_1$, server $S_2$ reconstructs the centered gradients as follows:
\begin{equation}
    \hat{\mathbf{g}}_i=\hat{\mathbf{g}}_i^{(1)}+\hat{\mathbf{g}}_i^{(2)}
\end{equation}

where random masks can cancel each other out:
\begin{equation}
    \begin{aligned}
        \hat{\mathbf{g}}_i & =\left(\tilde{\mathbf{g}}_i^{(1)}-\overline{\tilde{\mathbf{g}}}^{(1)}\right)+\left(\tilde{\mathbf{g}}_i^{(2)}-\overline{\tilde{\mathbf{g}}}^{(2)}\right) \\
        & =\left(\mathbf{g}_i^{(1)}+\mathbf{r}_i-\overline{\mathbf{g}}^{(1)}-\overline{\mathbf{r}}\right)+\left(\mathbf{g}_i^{(2)}-\mathbf{r}_i-\overline{\mathbf{g}}^{(2)}+\overline{\mathbf{r}}\right) \\
        & =\mathbf{g}_i^{(1)}+\mathbf{g}_i^{(2)}-\overline{\mathbf{g}}^{(1)}-\overline{\mathbf{g}}^{(2)} \\
        & =\mathbf{g}_i-\overline{\mathbf{g}}
    \end{aligned}
\end{equation}
Here, $\overline{\mathbf{g}}=\frac{1}{N} \sum_{j=1}^N \mathbf{g}_j$ denotes the global mean of the true (unmasked) gradients.

\begin{algorithm}
\caption{Gradient Mean-Central}\label{alg:MeanCentral}
\Input{Masked shares:$\tilde{\mathbf{g}}_i^{(1)}$ and 
$\tilde{\mathbf{g}}_i^{(2)}$}
\Output{$\hat{\mathbf{g}}_i$}

\Comment{Mean-Centering of Shares}
\textbf{Server $\mathcal{S}_1$: Mean-Centering of First Shares}\;
Receive $\tilde{\mathbf{g}}_i^{(1)}$ from all clients\;
\For{each edge device $i$}
{
 Compute:$\hat{\mathbf{g}}_i^{(1)}=\tilde{\mathbf{g}}_i^{(1)}-\frac{1}{N} \sum_{j=1}^N \tilde{\mathbf{g}}_j^{(1)}$ \;
}

Send $\hat{\mathbf{g}}_i^{(1)}$ to $\mathcal{S}_2$\;

\textbf{Server $\mathcal{S}_2$: Construction of Centered Gradient}
Receive $\hat{\mathbf{g}}_i^{(1)}$ from $\mathcal{S}_1$ and $\tilde{\mathbf{g}}_i^{(2)}$ from all clients\;

\For{each edge device $i$}
{
     Compute:$\hat{\mathbf{g}}_i^{(2)}=\tilde{\mathbf{g}}_i^{(2)}-\frac{1}{N} \sum_{j=1}^N \tilde{\mathbf{g}}_j^{(2)}$ \;
}
Obtain centered gradient:
$\hat{\mathbf{g}}_i = \hat{\mathbf{g}}_i^{(1)} + \hat{\mathbf{g}}_i^{(2)} = \mathbf{g}_i - \bar{\mathbf{g}}$\;
Use $\{\hat{\mathbf{g}}_i\}_{i=1}^N$ for following anomaly detection\;
\end{algorithm}

3) Malicious Gradient Detection: At this stage, server $\mathcal{S}_2$ conducts a hybrid anomaly detection procedure in collaboration with $\mathcal{S}_1$. The process integrates spectral analysis and cosine similarity to the reliability of client gradients.

\textbf{Spectral Projections:} Server $\mathcal{S}_2$ aggregates all centered gradients into a matrix:
\begin{equation}
    \mathbf{G}=\left[\hat{\mathbf{g}}_1, \hat{\mathbf{g}}_2, \ldots, \hat{\mathbf{g}}_N\right] \in \mathbb{R}^{d \times N}
\end{equation}
where each column corresponds to a centered gradient vector from a client. Singular value decomposition (SVD) \cite{hoecker1996svd} is then applied to extract the principal components:
\begin{equation}
    \mathbf{G}=\mathbf{U} \boldsymbol{\Sigma} \mathbf{V}^{\top}
\end{equation}
Here, $\mathbf{U} \in \mathbb{R}^{d \times d}$ contains the left singular vectors, $\boldsymbol{\Sigma} \in \mathbb{R}^{d \times N}$ is a diagonal matrix of singular values sorted in descending order, and $\mathbf{V} \in \mathbb{R}^{N \times N}$ contains the right singular vectors. The top right singular vector $\mathbf{v}_1$ (corresponding to the largest singular value) is selected as the dominant direction. Each client's gradient is projected onto $\mathbf{v}_1$ to compute its spectral score:
\begin{equation}
    s_i=\left(\hat{\mathbf{g}}_i^{\top} \mathbf{v}_1\right)^2
\end{equation}

Clients with high $s_i$ values are more likely to be malicious as their gradients deviate significantly from the principal component. 

\textbf{Cosine similarity:} To further assess the directional consistency,  the cosine similarity between each client’s centered gradient and others is calculated as: 
\begin{equation}
    c_{i j}=\frac{\hat{\mathbf{g}}_i^{\top} \hat{\mathbf{g}}_j}{\left\|\hat{\mathbf{g}}_i\right\| \cdot\left\|\hat{\mathbf{g}}_j\right\|}, \quad \forall i \neq j
\end{equation}

The median cosine similarity for each client is used as a reliability score:
\begin{equation}
    c_i=\operatorname{median}\left(\left\{c_{i j} \mid j \neq i\right\}\right)
\end{equation}

\textbf{Clustering-Based Detection:} Next, a two-dimensional feature vector is constructed for each client:
\begin{equation}
    \mathbf{f}_i=\left[s_i, c_i\right]^{\top}
\end{equation}
A K-means clustering algorithm with $K=2$ is applied to these features, producing two clusters $\mathcal{C}_1$ and $\mathcal{C}_2$ with centroids $\boldsymbol{\mu}_1$ and $\boldsymbol{\mu}_2$, respectively. To ensure reliable detection, a cluster separation check is performed. We compute the Euclidean distance between the two centroids:
\begin{equation}
    D=\left\|\mu_1-\mu_2\right\|
\end{equation}
and compare it with an adaptive threshold $\epsilon$ derived from the intra-cluster distances:
\begin{equation}
    \epsilon=\alpha \cdot \frac{1}{2}\left(\bar{d}_1+\bar{d}_2\right)
\end{equation}
where $\bar{d}_1=\frac{1}{\left|\mathcal{C}_1\right|} \sum_{\mathbf{f}_j \in \mathcal{C}_1}\left\|\mathbf{f}_j-\boldsymbol{\mu}_1\right\|$ and $\bar{d}_2=\frac{1}{\left|\mathcal{C}_2\right|} \sum_{\mathbf{f}_j \in \mathcal{C}_2}\left\|\mathbf{f}_j-\boldsymbol{\mu}_2\right\|$ denote the average intra-cluster distances, and $\alpha>1$ is a scaling factor. If $D<\epsilon$, the inter-cluster separation is smaller than the intra-cluster spread, indicating that all clients exhibit similar behavior in the current round. In this case, all clients are treated as benign and included in aggregation. If $D \geq \epsilon$, clients in the larger cluster are considered benign, while those in the smaller cluster are identified as anomalous and excluded from aggregation. This adaptive separation check avoids false exclusion of benign clients in attack-free rounds while maintaining effective detection when attacks are present.
% A K-means clustering algorithm \cite{liu10143180} (with a predefined number of clusters $K=2$) is applied to these features. Clients in the largest cluster are considered benign, while those in smaller clusters are viewed as anomalous and excluded from aggregation. This approach allows efficient detection of malicious behaviors while preserving gradient privacy.
\begin{algorithm}[t]
\caption{Privacy-Preserving Malicious Gradient Detection}
\label{alg:byzantine_detection}
\KwIn{Mean-centered masked gradients $\hat{\mathbf{g}}_i^{(1)}$ from $\mathcal{S}_1$, masked gradients $\tilde{\mathbf{g}}_i^{(2)}$ from clients}
\KwOut{Trusted gradient set $\mathcal{G}$}

Stack all $\hat{\mathbf{g}}_i$ into matrix $G \in \mathbb{R}^{N \times d}$

Perform SVD: $G = U \Sigma V^\top$; let $v_1$ be the top right singular vector

\ForEach{client $i$}{
    Compute spectral score: $s_i^{\text{svd}} \leftarrow |\langle \hat{\mathbf{g}}_i, v_1 \rangle|$ \\
    Compute cosine similarity with others: \\
    \quad $c_i \leftarrow \operatorname{median}(\{ \cos(\hat{\mathbf{g}}_i, \hat{\mathbf{g}}_j) \}_{j \neq i})$ \\
    Construct feature vector: $\phi_i \leftarrow [s_i^{\text{svd}}, c_i]$
}

Apply Mean Shift clustering on $\{ \phi_i \}_{i=1}^N$

Let $\mathcal{G} \leftarrow$ the cluster with the largest number of clients

\Return{$\mathcal{G}$}
\end{algorithm}

4) Trust Scoring-Based Weight Computation: The trust scoring mechanism ensures that individual client nodes contribute appropriately to the global model aggregation process \cite{tong2024multi}. Unlike prior approaches that rely solely on per-round behavior, DP2Guard introduces a time-evolving trust evaluation scheme that captures clients' long-term reliability.

In each round $t$, server $\mathcal{S}_2$ calculates the Euclidean distance between each client's feature vector $\mathbf{f}_i^{(t)}$ and the benign cluster centroid $\boldsymbol{\mu}^{(t)}$:
\begin{equation}
    \operatorname{Dis}_i^{(t)}=\left\|\mathbf{f}_i^{(t)}-\boldsymbol{\mu}^{(t)}\right\|
\end{equation}
Based on this distance, the direct trust score for device $E_i$ is computed as:
\begin{equation}
    \gamma_i^{(t)}=1 /\left(1+\operatorname{Dis}_i^{(t)}\right)
\end{equation}
For clients in the benign cluster, their feature vectors are close to the centroid, resulting in a small $\mathrm{Dis}_i{ }^{(t)}$ and a high $\gamma_i^{(t)}$. For clients assigned to the anomalous cluster, their feature vectors are far from the benign centroid, leading to a large $\operatorname{Dis}_i^{(t)}$ and a near-zero $\gamma_i^{(t)}$.

To balance historical performance and recent behavior, DP2Guard utilizes a time-based trust mechanism. For each client $i$, the trust score is updated as:
\begin{equation}
    \operatorname{Trust}_i^{(t)}=\beta \cdot \operatorname{Trust}_i^{(t-1)}+(1-\beta) \cdot \gamma_i^{(t)}, \quad \beta \in[0,1)
\end{equation}
Note that this update is applied to all clients in each round, including those excluded by the clustering mechanism. For excluded clients, their near-zero $\gamma_i^{(t)}$ causes a significant drop in their trust score, ensuring that clients exhibiting malicious behavior are penalized even when their gradients are already excluded from aggregation. This is particularly effective against on-off attacks where adversaries alternate between benign and malicious behavior.
The normalized trust value is used to compute the aggregation weight:
$$
\tau_i^{(t)}=\frac{\operatorname{Trust}_i^{(t)}}{\sum_{j=1}^N \operatorname{Trust}_j^{(t)}}
$$
Server $\mathcal{S}_2$ then applies these weights to aggregate the received masked gradients:
\begin{equation}
    \tilde{\mathbf{g}}_{\text {agg }}^{(2)}=\sum_{i=1}^N \tau_i^{(t)} \cdot \tilde{\mathbf{g}}_i^{(2)}
\end{equation}
Then, both $\tilde{\mathbf{g}}_{\text {agg }}^{(2)}$ and trust weights $\left\{\tau_i^{(t)}\right\}_{i=1}^N$ are published to the blockchain. 

\textbf{4) Adaptive Aggregation:} In the stage, server $\mathcal{S}_1$ downloads the aggregated masked gradient $\tilde{\mathbf{g}}_{\text {agg }}^{(2)}$ and corresponding trust scores from blockchain. Using the same trust weights, $\mathcal{S}_1$ performs weighted aggregation over its local set of masked gradients:
\begin{equation}
    \tilde{\mathbf{g}}_{\mathrm{agg}}^{(1)}=\sum_{i=1}^N \tau_i^{(t)} \cdot \tilde{\mathbf{g}}_i^{(1)}
\end{equation}
The final gradient can be reconstructed by combining the two:
\begin{equation}
\begin{aligned}
\tilde{\mathbf{g}}_{\text{agg}} &= \tilde{\mathbf{g}}_{\text{agg}}^{(1)} + \tilde{\mathbf{g}}_{\text{agg}}^{(2)} \\
&= \sum_{i=1}^N \tau_i^{(t)} \left( \mathbf{g}_i^{(1)} + \mathbf{r}_i + \mathbf{g}_i^{(2)} - \mathbf{r}_i \right) \\
&= \sum_{i=1}^N \tau_i^{(t)} \left( \mathbf{g}_i^{(1)} + \mathbf{g}_i^{(2)} \right) \\
&= \sum_{i=1}^N \tau_i^{(t)} \cdot \mathbf{g}_i
\end{aligned}
\end{equation}
This aggregated result $\tilde{\mathbf{g}}_{\text {agg }}$ is then submitted to the blockchain to support model updates in the next training round.

\begin{algorithm}[ht]
\caption{Trust-Weighted Adaptive Aggregation}
\label{alg:trust_weighted_aggregation}
\KwIn{Trust weights $\{\tau_i^{(t)}\}_{i=1}^N$,\\
\hspace{1.1cm} Masked gradients $\{\tilde{\mathbf{g}}_i^{(1)}\}_{i=1}^N$ at $\mathcal{S}_1$, $\{\tilde{\mathbf{g}}_i^{(2)}\}_{i=1}^N$ at $\mathcal{S}_2$}
\KwOut{Global aggregated gradient $\tilde{\mathbf{g}}_{\text{agg}}$}

\textbf{At Server $\mathcal{S}_2$:} \\
\quad Compute weighted aggregation of second shares: \\
\quad $\tilde{\mathbf{g}}_{\text{agg}}^{(2)} \leftarrow \sum_{i=1}^N \tau_i^{(t)} \cdot \tilde{\mathbf{g}}_i^{(2)}$ \\
\quad Send $\tilde{\mathbf{g}}_{\text{agg}}^{(2)}$ and $\{\tau_i^{(t)}\}$ to blockchain \\[0.3em]

\textbf{At Server $\mathcal{S}_1$:} \\
\quad Download the $\tilde{\mathbf{g}}_{\text{agg}}^{(2)}$ and $\{\tau_i^{(t)}\}$ from blockchain \\
\quad Compute weighted aggregation of first shares: \\
\quad $\tilde{\mathbf{g}}_{\text{agg}}^{(1)} \leftarrow \sum_{i=1}^N \tau_i^{(t)} \cdot \tilde{\mathbf{g}}_i^{(1)}$ \\
\quad Reconstruct the global aggregated gradient: \\
\quad $\tilde{\mathbf{g}}_{\text{agg}} \leftarrow \tilde{\mathbf{g}}_{\text{agg}}^{(1)} + \tilde{\mathbf{g}}_{\text{agg}}^{(2)}$ \\

\Return{$\tilde{\mathbf{g}}_{\text{agg}}$}
\end{algorithm}

\section{Security Analysis}
\label{sec:Security} 
In this section, we prove the privacy and robustness of DP2Guard, and analyze the computational and communication complexity in DP2Guard.

\begin{theorem}[Privacy protection against honest-but-curious servers]
Under the assumption that servers $S_1$ and $S_2$ do not collude, the execution of DP2Guard does not leak any information about the true local gradients $\mathbf{g}_i$ to either server individually.
\end{theorem}\label{th:privacy}

We use a standard hybrid argument \cite{lindell2017simulate} to prove the Theorem 1. Let ${REAL}_{\mathcal{A}}^{\Pi}$ denote the view of an adversary $\mathcal{A}$ during the real execution of protocol $\Pi$, and let ${IDEAL}_{\mathcal{A}^*}^{\mathcal{F}}$ denote the ideal execution where a trusted functionality $\mathcal{F}$ performs all privacy-sensitive operations. We construct a sequence of hybrid experiments, $H_0, H_1, \ldots, H_5$, such that each pair of consecutive hybrids is computationally indistinguishable.

$\textbf{Hyb}_0$: This hybrid corresponds to the actual execution of the DP2Guard protocol (as described in Section \ref{sec:OurScheme}). Each client computes its true local gradient $\mathrm{g}_i^{(t)}$, splits it into two parts $\mathrm{g}_i^{(1)}$ and $\mathrm{g}_i^{(2)}$, and applies a random mask $\mathbf{r}_i$ to obtain $\tilde{\mathrm{g}}_i^{(1)}=\mathrm{g}_i^{(1)}+\mathrm{r}_i$ and $\tilde{\mathrm{g}}_i^{(2)}=\mathrm{g}_i^{(2)}-\mathrm{r}_i$, which are sent to $\mathcal{S}_1$ and $\mathcal{S}_2$, respectively.

$\textbf{Hyb}_1$: In this hybrid, the simulator replaces the masked gradient share $\tilde{\mathbf{g}}_i^{(1)}$ sent to $\mathcal{S}_1$ with a uniformly random vector $\chi_i^{(1)}$ of the same dimension. The second share $\tilde{\mathbf{g}}_i^{(2)}$ remains unchanged. Since $\chi_i^{(1)}$ is independent and indistinguishable from a valid masked share, and $\mathcal{S}_1$ does not access the corresponding second share, this view is computationally indistinguishable from $\mathbf{Hyb}_0$.

{$\textbf{Hyb}_2$}: Building on $\mathbf{Hyb}_1$, this hybrid replaces the share sent to $\mathcal{S}_2$ as well. Each client sends a uniformly random vector $\chi_i^{(2)}$ to $\mathcal{S}_2$, where $\chi_i^{(1)} + \chi_i^{(2)} = \chi_i$ and $\chi_i$ is sampled uniformly at random. As both shares are now random and the servers are non-colluding, this hybrid is indistinguishable from $\mathbf{Hyb}_1$.

{$\textbf{Hyb}_3$}: In this hybrid, we simulate the internal computations of $\mathcal{S}_1$ that are transmitted to $\mathcal{S}_2$, such as the mean-centered gradients $\hat{\mathbf{g}}_i^{(1)}$. Specifically, for each client $i$, the simulator replaces $\hat{\mathbf{g}}_i^{(1)}$ with a randomly generated vector $\psi_i^{(1)}$ sampled from a distribution that is statistically consistent with that of the true centered values. Since $\mathcal{S}_2$ does not observe the raw gradients or the corresponding masked shares held by $\mathcal{S}_1$, its view remains computationally indistinguishable from $\mathbf{Hyb}_2$.

{$\textbf{Hyb}_4$}: Similarly, this hybrid simulates the internal computations of $\mathcal{S}_2$ that are sent to $\mathcal{S}_1$ (e.g., trust scores or aggregation results). Each intermediate value is replaced by a random vector $\psi_i^{(2)}$ drawn from a consistent distribution. As $\mathcal{S}_1$ does not observe the raw gradients or masked shares held by $\mathcal{S}_2$, its view remains indistinguishable from $\mathbf{Hyb}_3$.

{$\textbf{Hyb}_5$}: Finally, we simulate the messages that $\mathcal{S}_1$ sends back to the clients (e.g., global model updates) using random values that are statistically indistinguishable from actual aggregated updates. This completes the transition from the real-world protocol to an ideal-world simulation.

By the transitivity of computational indistinguishability, we conclude that the output of the simulator $\mathsf{SIM}$ is indistinguishable from the real execution:
$$
\mathrm{REAL}_{\mathcal{A}}^{\Pi} \approx \operatorname{IDEAL}_{\mathcal{A}^*}^{\mathcal{F}}
$$

Hence, DP2Guard preserves the privacy of local gradients under the honest-but-curious model, ensuring that neither $\mathcal{S}_1$ nor $\mathcal{S}_2$ can learn any sensitive information individually.

\section{Performance Evaluation}
\label{sec:Result}

\subsection{Experimental Settings}

\begin{figure}[t]
\centering
\begin{subfigure}[t]{0.48\columnwidth}
\centering
\includegraphics[width=\linewidth]{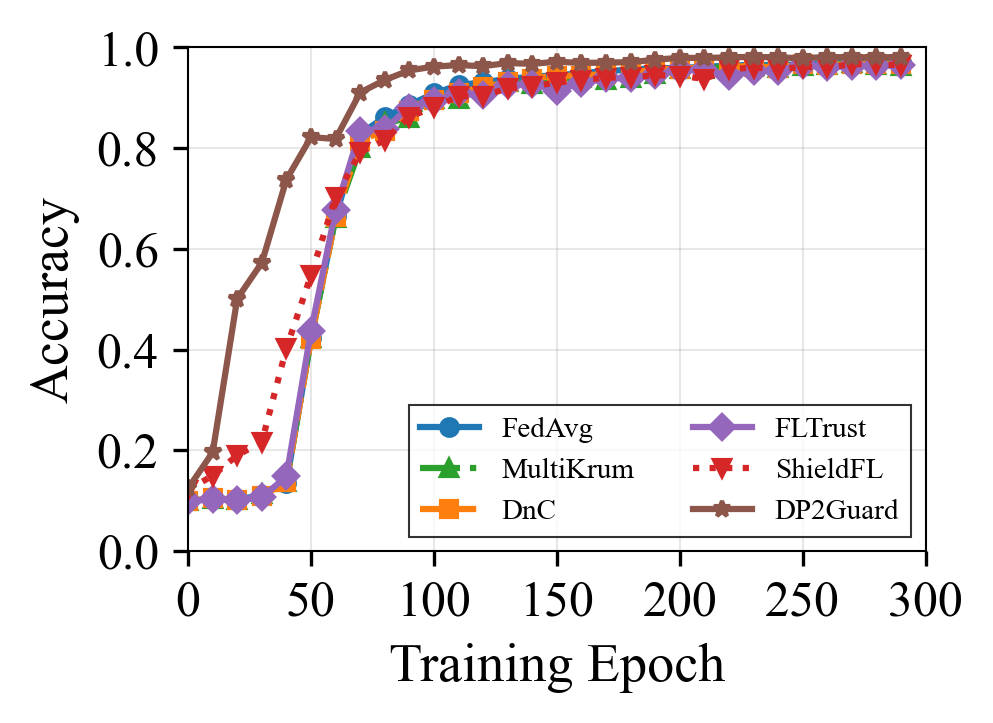}
\caption{MNIST (IID)}
\label{fig:perf_a}
\end{subfigure}
\hfill
\begin{subfigure}[t]{0.48\columnwidth}
\centering
\includegraphics[width=\linewidth]{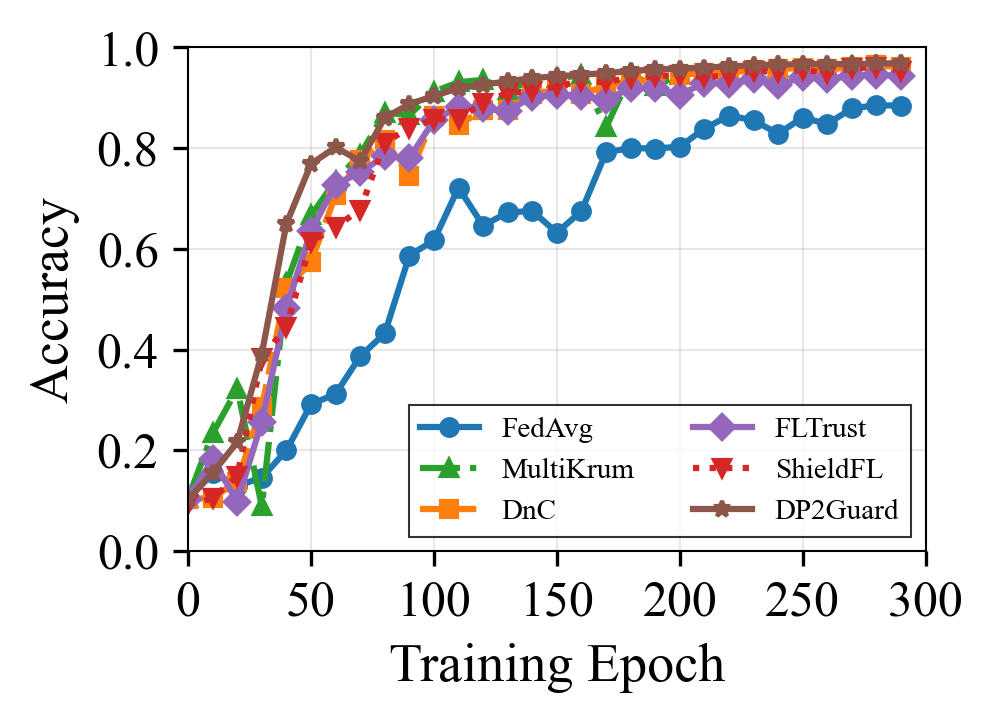}
\caption{MNIST (Non-IID)}
\label{fig:perf_b}
\end{subfigure}

\vspace{2mm}

\begin{subfigure}[t]{0.48\columnwidth}
\centering
\includegraphics[width=\linewidth]{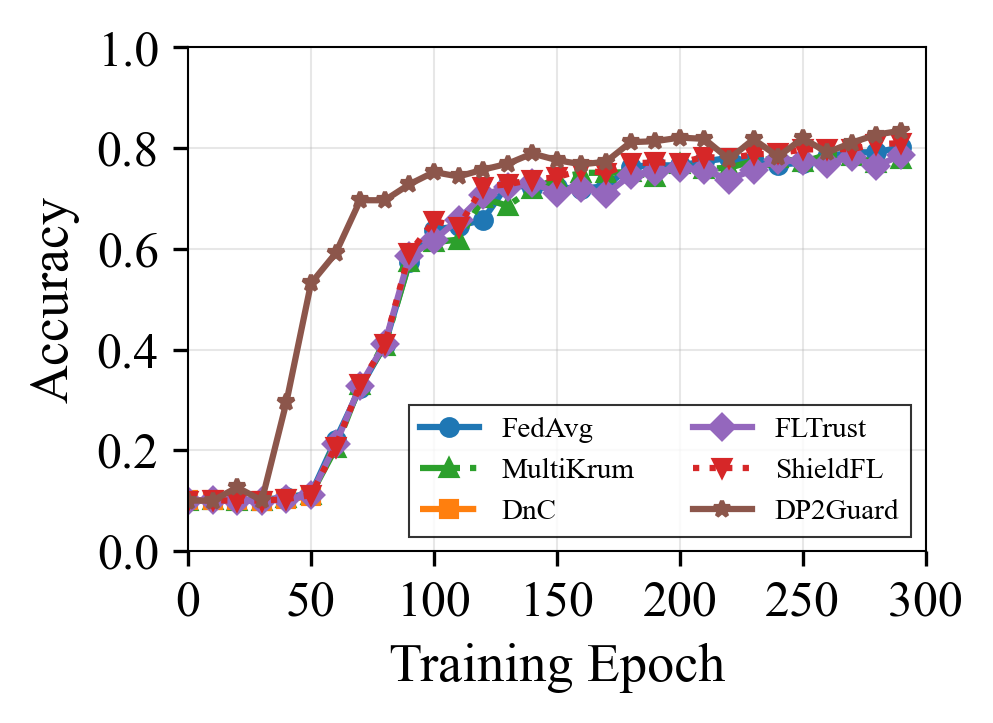}
\caption{Fashion-MNIST (IID)}
\label{fig:perf_c}
\end{subfigure}
\hfill
\begin{subfigure}[t]{0.48\columnwidth}
\centering
\includegraphics[width=\linewidth]{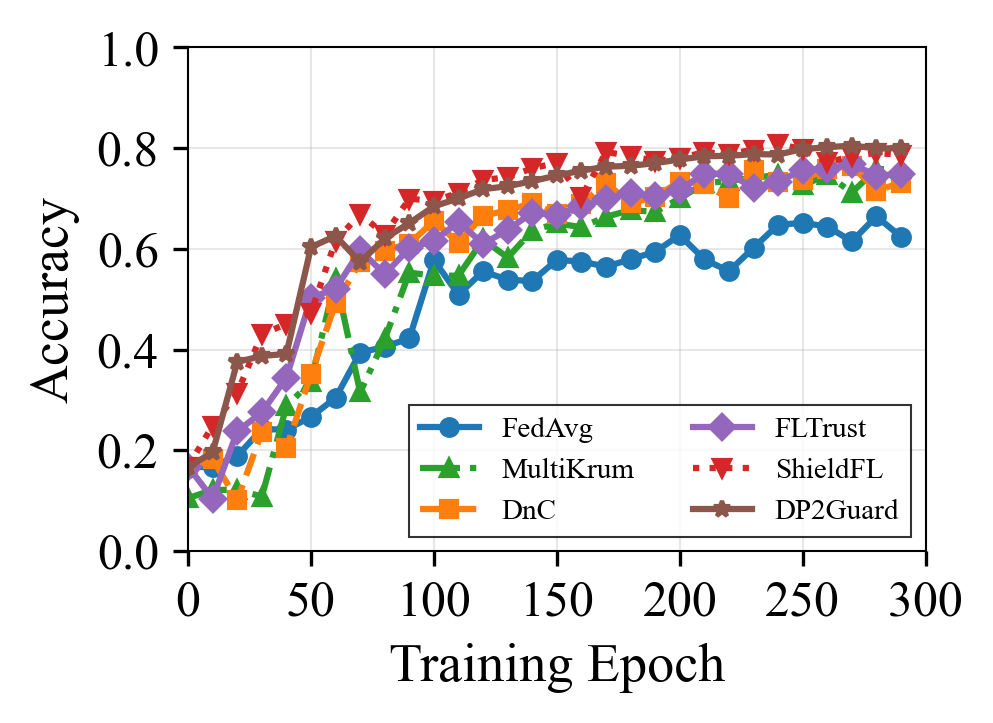}
\caption{Fashion-MNIST (Non-IID)}
\label{fig:perf_d}
\end{subfigure}

\vspace{2mm}

\begin{subfigure}[t]{0.48\columnwidth}
\centering
\includegraphics[width=\linewidth]{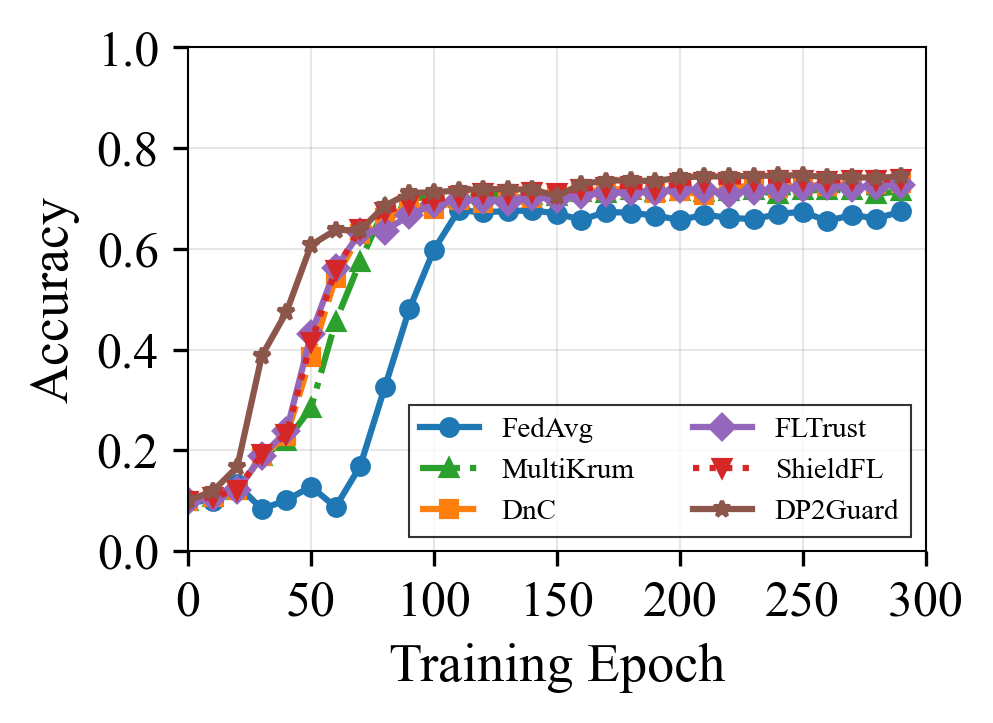}
\caption{CIFAR10 (IID)}
\label{fig:perf_c}
\end{subfigure}
\hfill
\begin{subfigure}[t]{0.48\columnwidth}
\centering
\includegraphics[width=\linewidth]{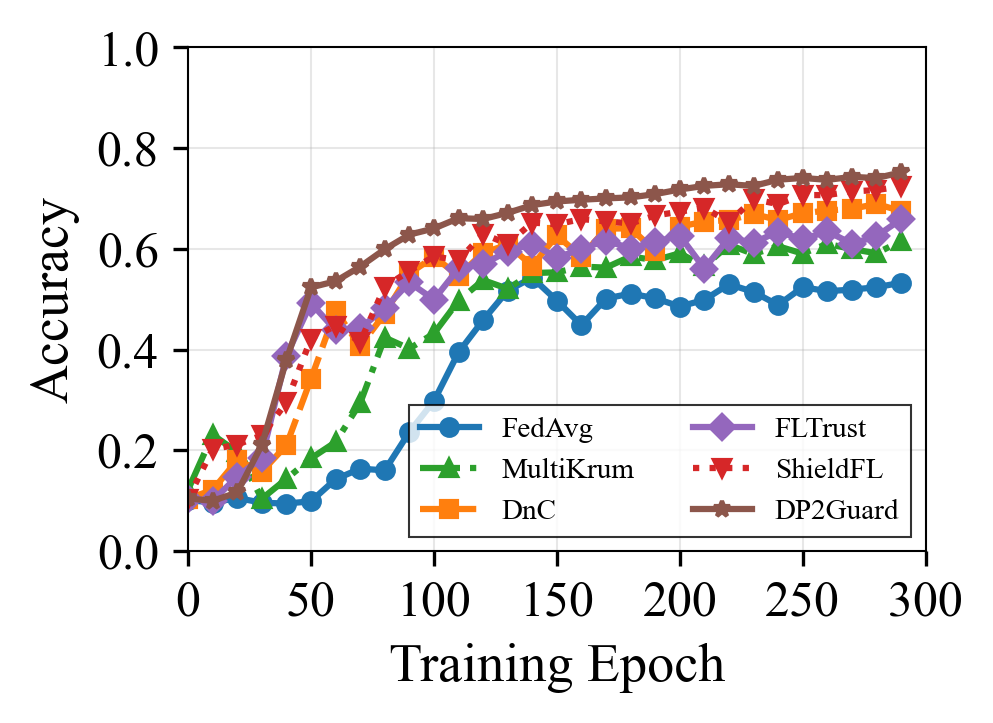}
\caption{CIFAR10 (Non-IID)}
\label{fig:perf_d}
\end{subfigure}

\caption{Impact of training iterations on the effectiveness of defense strategies on MNIST and Fashion-MNIST datasets under IID and Non-IID settings.}
\label{fig:performance}

\end{figure}

% \begin{figure*}[t]
%   \centering
%   \begin{subfigure}[b]{0.24\textwidth}
%     \includegraphics[width=\linewidth]{result/noattackiid01.png}
%     \caption{MNIST IID}
%   \end{subfigure}
%   \hfill
%   \begin{subfigure}[b]{0.24\textwidth}
%     \includegraphics[width=\linewidth]{result/noattackiid_fashion.png}
%     \caption{Fashion IID}
%   \end{subfigure}
%   \hfill
%   \begin{subfigure}[b]{0.24\textwidth}
%     \includegraphics[width=\linewidth]{result/noattackmnistnoiid.png}
%     \caption{MNIST Non-IID}
%   \end{subfigure}
%   \hfill
%   \begin{subfigure}[b]{0.24\textwidth}
%     \includegraphics[width=\linewidth]{result/noattacknoniid_fashion.png}
%     \caption{Fashion Non-IID}
%   \end{subfigure}
  
%   \caption{Impact of training iterations on the effectiveness of defense strategies on MNIST and Fashion-MNIST datasets under IID and Non-IID settings.}
%   \label{fig:performance}
% \end{figure*}

All experiments were conducted on a high-performance workstation running Ubuntu 20.04 LTS, equipped with an Intel i9 CPU, 64 GB of RAM, and four NVIDIA RTX 4090 GPUs. The FL framework was implemented using PyTorch version 1.6.0 and Python version 3.8.10.

\textbf{Datasets and Model Architectures:} We evaluate our proposed scheme on three benchmark datasets: MNIST, Fashion-MNIST, and CIFAR-10. 
The MNIST dataset contains 60,000 training and 10,000 testing grayscale images of handwritten digits, each of size $28 \times 28$ pixels and classified into 10 categories. 
The Fashion-MNIST dataset consists of 60,000 training and 10,000 testing grayscale images representing various types of clothing items, also distributed across 10 classes. 
The CIFAR-10 dataset contains 50,000 training and 10,000 testing color images of size $32 \times 32$ pixels across 10 classes. 
For MNIST and Fashion-MNIST, we employ the LeNet5 model \cite{lecun2002gradient} to perform the training tasks.
For CIFAR-10, we adopt the deeper ResNet18 model \cite{he2016deep} to better handle the increased complexity of the color image dataset. For all experiments,
we adopt stochastic gradient descent (SGD) as the optimizer, with a learning rate $\eta=0.01$ and a batch size of 32.

\begin{figure}
    \centering
    \includegraphics[width=1.0\linewidth]{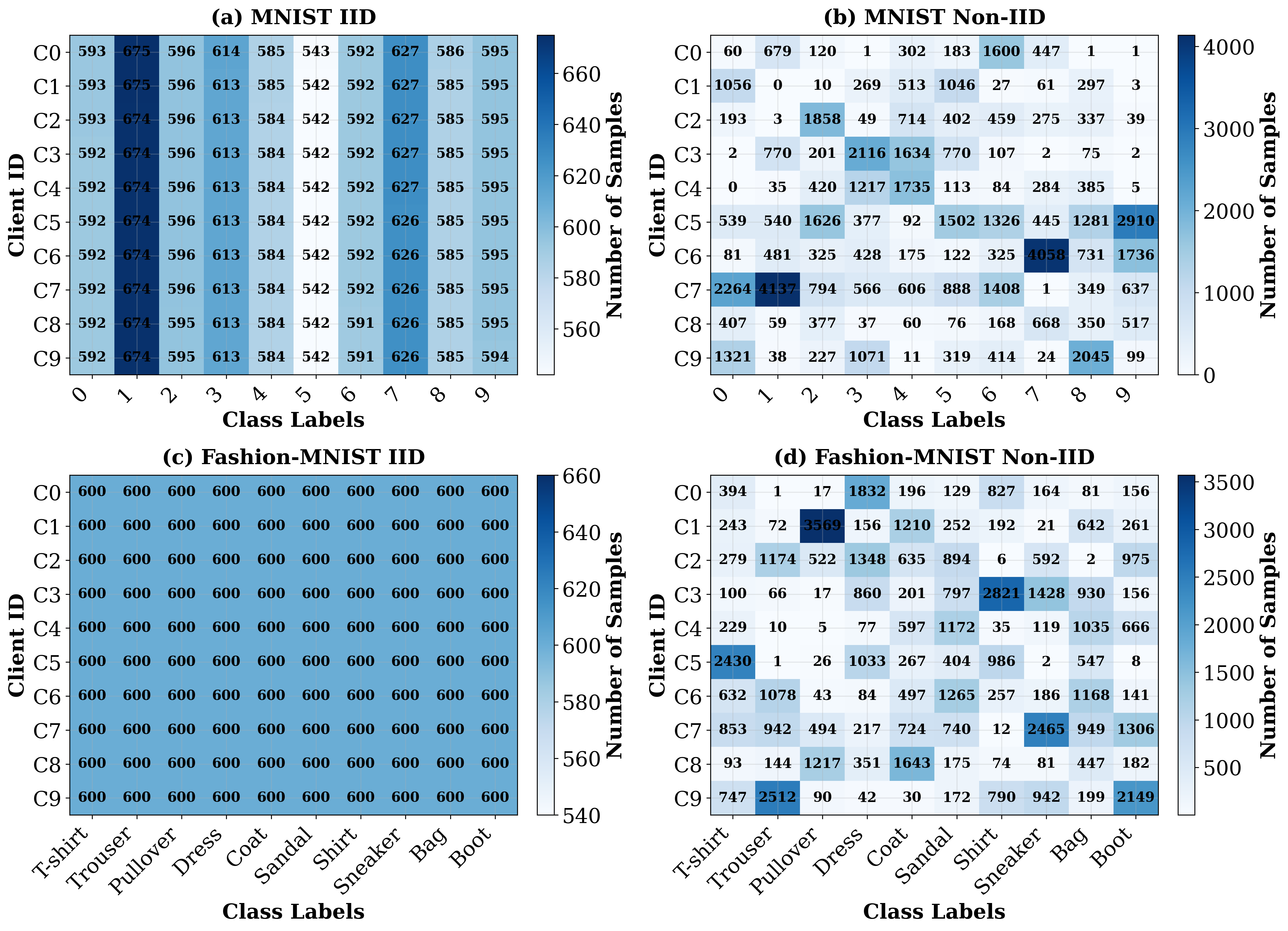}
    \caption{Class-wise sample distribution heatmap for 10 randomly selected clients in a FL setup with 50 clients. The datasets used are FashionMNIST and MNIST. Data partitioning is performed using a Dirichlet distribution with a heterogeneity parameter of $\alpha=0.5$ to simulate non-IID conditions.}
    \label{fig:partitioningscheme}
\end{figure}

\textbf{Datasets Distribution:} We conduct experiments under both independent and identically distributed (IID) and non-IID data settings. In the IID setting, data samples are randomly and uniformly distributed among clients. For the non-IID setting, we adopt the commonly used Dirichlet distribution strategy \cite{Chen10787234} to partition the datasets among clients. According to the configuration in prior work \cite{bagdasaryan2020backdoor}, the data heterogeneity parameter is set to $\alpha=0.5$, where a smaller $\alpha$ indicates higher data skewness and greater class imbalance across clients. Fig. \ref{fig:partitioningscheme} illustrates the class distribution of $10$ selected clients out of a total of 50 under this partitioning scheme.

\textbf{FL System Settings:} In our experiments, we set up 50 clients for FL training process. Each client computes local gradients and sends them to the central server, which aggregates these updates using different aggregation rules. The FedAvg \cite{McMahan} is adopted as the default aggregation mechanism. The training is conducted over $300$ global communication rounds, with each client performing one local epoch per round. To assess the impact of adversarial behavior, we set the ratio of adversary Att $_{\text {ratio }}=\{0 \%, 10 \%, 20 \%, 30 \%, 40 \%\}$.

\textbf{Baselines:} We compare DP2Guard with the following state-of-the-art methods. 
For robust FL, we consider Multi-Krum \cite{Blanchard}, DnC \cite{shejwalkar2021manipulating}, and FLTrust \cite{cao2020fltrust}, which focus on defending against poisoning attacks. 
For privacy-preserving FL, we include PBFL \cite{Ren9846908}, which employs a double-masking protocol to protect gradient privacy but does not support Byzantine attack detection. For robust privacy-preserving FL, we consider ShieldFL \cite{MaZhuoran}, which integrates cosine similarity-based detection with a two-trapdoor HE protocol.
These baselines have been discussed in detail in Section \ref{sec:relatedwork}. 
Note that since PBFL does not incorporate Byzantine-resilient aggregation, it is excluded from the robustness evaluation against poisoning attacks and is only included in the privacy and efficiency comparisons.

\textbf{Attack Type:} To comprehensively evaluate the robustness of DP2Guard, we consider four advanced poisoning attack scenarios:

\begin{enumerate}
    \item \textbf{Label-Flipping attack:} Label-flipping attacks \cite{biggio2012poisoning} manipulate the labels of selected training samples on the malicious client, causing the local model to learn incorrect class boundaries. For example, samples with the true label $2$ may be maliciously relabeled as $7$ to cause targeted misclassification. In our experiments, each malicious client flips 30\% of its local labels.
        
    \item \textbf{Fang Attack:} In this adaptive attack, the adversary is assumed to have full knowledge of all benign gradients and the aggregation rule. Following the setting in \cite{fang2020local}, we set the perturbation threshold $\gamma_{\min}=1 e-5$, and the optimization step size to $\alpha=0.5$. The unit perturbation direction is determined by the sign of the pre-aggregation gradient computed locally by the adversary.
    
    \item \textbf{Min-Max \cite{shejwalkar2021manipulating}:} In this adaptive attack, the adversary has access to all benign client gradients but does not know the aggregation rule. An efficient binary search algorithm is used to determine the optimal perturbation coefficient $\lambda$. The initial perturbation factor is set to $\gamma=10$, with an initial step size of $\gamma / 2=5$, and the search stops when the perturbation threshold $\gamma_{\min }=1 \times 10^{-5}$ is reached. The direction of the perturbation is defined as the unit vector of the mean of the known benign gradients.
    
    \item \textbf{Min-Sum Attack \cite{shejwalkar2021manipulating}:} Min-Sum follows the same procedure as Min-Max but changes the objective: instead of maximizing the largest deviation, it minimizes the sum of squared distances to all benign gradients. A binary search is used with $\gamma=10$, step size $=5$, and stopping threshold $\gamma_{\min }=1 \times 10^{-5}$. The perturbation direction is the unit vector of the mean benign gradient.
\end{enumerate}

\newcommand{\best}[1]{\cellcolor{gray!25}\textbf{#1}}
\newcommand{\sbest}[1]{\cellcolor{gray!10}{#1}}
\begin{table*}[t]
    \centering
    \small
    \setlength{\tabcolsep}{4.5pt}
    \renewcommand{\arraystretch}{1.35}
    \caption{Comparison of defense strategies against four poisoning attacks on MNIST, Fashion-MNIST, and CIFAR-10 in the IID setting, with malicious-client ratios ranging from 10\% to 40\%. The \best{best} and \sbest{second-best} results are highlighted.}
    \label{tab:def_per_iid}
    
    \begin{tabular}{ll cccc cccc cccc cccc}
    \toprule
    \multirow{2}{*}{\textbf{Data set}} & \multirow{2}{*}{\textbf{Method}}
    & \multicolumn{4}{c}{\textbf{LF Attack}}
    & \multicolumn{4}{c}{\textbf{Fang Attack}}
    & \multicolumn{4}{c}{\textbf{Min-Max}}
    & \multicolumn{4}{c}{\textbf{Min-Sum Attack}} \\
    \cmidrule(lr){3-6}\cmidrule(lr){7-10}\cmidrule(lr){11-14}\cmidrule(lr){15-18}
    
    &
    & \textbf{10} & \textbf{20} & \textbf{30} & \textbf{40}
    & \textbf{10} & \textbf{20} & \textbf{30} & \textbf{40}
    & \textbf{10} & \textbf{20} & \textbf{30} & \textbf{40}
    & \textbf{10} & \textbf{20} & \textbf{30} & \textbf{40}\\
    \midrule
    
    \multirow{6}{*}{\textbf{MNIST}}
    & FedAvg     & 95.2 & 93.4 & 90.1 & 85.3  & 90.0 & 84.2 & 65.4 & 10.1 & 95.0 & 92.4 & 87.0 & 80.3  & 94.0 & 90.4 & 86.0 & 82.9 \\
    & FLTrust    & 96.5 & 96.2 & 95.9 & 94.6 & \best{96.4} & \sbest{95.0} & 94.0 & \sbest{93.4}  & 95.9 & 95.2 & 94.7 & 94.1  & 96.0 & 95.4 & 95.0 & 94.7 \\
    & MultiKrum  & 96.0 & 95.8 & 95.4 & 94.0  & 95.1 & 90.8 & 85.2 & 74.3  & 96.2 & 95.0 & 94.0 & 85.5  & 95.7 & 94.0 & 92.0 & 86.3 \\
    & ShieldFL  & 96.9 & \sbest{96.6} & 96.24 & 94.5 & \sbest{96.0} & \best{95.4} & \sbest{94.1} & \sbest{93.3} & \sbest{96.8} & \sbest{96.0} & \sbest{95.2} & \sbest{94.4} & \sbest{96.5} & \sbest{96.0} & \sbest{95.6} & \sbest{95.1} \\
    & DnC     & \sbest{97.0} & 96.59 & \sbest{96.26} & \sbest{95.0} & 95.1 & 93.6 & 93.0 & 92.5 & 96.0 & 95.4 & 95.0 & 94.1 & 95.9 & 95.2 & 94.7 & 93.7 \\
    & DP2Guard   & \best{97.2} & \best{96.9} & \best{96.6} & \best{95.4} & \best{96.4} & \best{95.4} & \best{95.3} & \best{95.0} & \best{97.0} & \best{96.5} & \best{96.0} & \best{95.2}  & \best{97.0} & \best{96.5} & \best{96.0} & \best{95.7} \\
    \midrule

    \multirow{6}{*}{\textbf{Fashion-MNIST}}
    & FedAvg     &78.0 & 77.4 & 76.0 & 72.3 & 72.0 & 65.4 & 50.0 & 10.0  & 76.0 & 74.4 & 69.0 & 61.1  & 76.0 & 72.4 & 67.0 & 61.1  \\
    & FLTrust    & 79.2 & \sbest{79.0} & 78.5 & 78.2 & 79.2 & \sbest{79.0} & \sbest{78.5} & \sbest{74.7}   & 79.2 & \sbest{79.0} & \sbest{78.5} & 74.7  & 79.0 & \sbest{78.7} & 78.0 & 75.7 \\
    & MultiKrum  & 78.7 & 78.4 & 78.2 & 78.0  & 75.7 & 72.4 & 63.2 & 47.2  & 77.7 & 76.4 & 74.2 & 68.6 & 77.7 & 75.4 & 72.2 & 68.6 \\
    & ShieldFL   & 79.3 & 78.6 & 78.4 & 78.3 & \sbest{79.3} & 78.6 & 78.4 & 74.2   & \sbest{79.3} & 78.6 & 78.4 & \sbest{76.9}  & \sbest{79.3} & 78.6 & \sbest{78.4} & \sbest{76.9} \\
    & DnC        & \sbest{79.5} & \sbest{79.0} & \sbest{78.7} & \sbest{78.4}  & 79.2 & 77.2 & 76.7 & 73.9  & 79.2 & 78.2 & 76.7 & 74.1   & 78.2 & 77.0 & 75.7 & 74.4 \\
    & DP2Guard   & \best{79.9} & \best{79.5} & \best{79.2} & \best{78.7}  & \best{79.9} & \best{79.5} & \best{79.4} & \best{78.2} & \best{79.9} & \best{79.5} & \best{79.4} & \best{78.6}  & \best{79.4} & \best{78.9} & \best{78.7} & \best{78.3}\\
    \midrule

    \multirow{6}{*}{\textbf{CIFAR-10}}
      & FedAvg     & 65.0 & 62.4 & 58.0 & 50.3  & 58.0 & 47.4 & 28.0 & 10.0  &
      64.0 & 61.4 & 55.0 & 44.1  & 64.0 & 60.4 & 55.0 & 47.1 \\
      & FLTrust    & 72.5 & 72.0 & 71.5 & 70.2  & 72.5 & 71.0 & \sbest{70.5} & \sbest{68.7}  &
      72.2 & 71.0 & 70.5 & 68.7  & 72.0 & 71.7 & 71.0 & 69.7 \\
      & MultiKrum  & 71.0 & 70.4 & 69.2 & 67.0  & 68.1 & 62.8 & 52.2 & 38.3  &
      70.2 & 68.0 & 64.2 & 57.6  & 70.7 & 68.0 & 64.2 & 57.6 \\
      & ShieldFL   & 73.0 & 72.6 & \sbest{72.4} & 71.5  & \sbest{73.0} & \sbest{71.4} & 70.1 & 67.2  &
      \sbest{73.0} & \sbest{72.0} & \sbest{71.2} & \sbest{70.4}  & \sbest{73.3} & \sbest{72.6} & \sbest{72.0} & \sbest{70.9} \\
      & DnC        & \sbest{73.5} & \sbest{72.9} & \sbest{72.4} & \sbest{71.8}  & 71.1 & 69.6 & 69.0 & 66.5  &
      72.0 & 71.4 & 71.0 & 69.1  & 72.2 & 71.2 & 70.7 & 69.4 \\
      & DP2Guard   & \best{74.2} & \best{73.9} & \best{73.5} & \best{72.4}  & \best{73.4} & \best{72.5} & \best{72.3} & \best{71.0}  &
      \best{74.0} & \best{73.5} & \best{73.0} & \best{72.2}  & \best{74.0} & \best{73.5} & \best{73.0} & \best{72.7} \\
    
    \bottomrule
\end{tabular}
\end{table*}

\begin{table*}[t]
\centering
\small
% \normalsize
\setlength{\tabcolsep}{4.5pt}
\renewcommand{\arraystretch}{1.35}
\caption{Comparison of defense strategies against four poisoning attacks on MNIST, Fashion-MNIST, and CIFAR-10 in the Non-IID setting, with malicious-client ratios ranging from 0\% to 40\%. The \best{best} and \sbest{second-best} results are highlighted.}
\label{tab:def_per_noniid}
\begin{tabular}{ll cccc cccc cccc cccc}
\toprule
\multirow{2}{*}{\textbf{Data set}} & \multirow{2}{*}{\textbf{Method}}
& \multicolumn{4}{c}{\textbf{LF Attack}}
& \multicolumn{4}{c}{\textbf{Fang Attack}}
& \multicolumn{4}{c}{\textbf{Min-Max}}
& \multicolumn{4}{c}{\textbf{Min-Sum Attack}} \\
\cmidrule(lr){3-6}\cmidrule(lr){7-10}\cmidrule(lr){11-14}\cmidrule(lr){15-18}

&
& \textbf{10} & \textbf{20} & \textbf{30} & \textbf{40}
& \textbf{10} & \textbf{20} & \textbf{30} & \textbf{40}
& \textbf{10} & \textbf{20} & \textbf{30} & \textbf{40}
& \textbf{10} & \textbf{20} & \textbf{30} & \textbf{40}\\
\midrule

\multirow{6}{*}{\textbf{MNIST}}
  & FedAvg     & 91.2 & 88.4 & 83.0 & 76.3  & 84.0 & 72.2 & 42.4 & 10.0  &
  90.0 & 86.4 & 79.0 & 69.3  & 89.0 & 84.4 & 78.0 & 72.9 \\
  & FLTrust    & 94.5 & 94.0 & 93.5 & \sbest{92.2}  & \sbest{94.0} & 92.5 & 91.0 & \sbest{89.4}  &
  94.0 & 93.0 & 92.0 & 90.1  & 94.0 & 93.4 & 92.5 & 90.7 \\
  & MultiKrum  & 93.0 & 92.4 & 91.0 & 87.0  & 92.1 & 82.8 & 70.2 & 53.3  &
  93.0 & 90.0 & 86.0 & 74.5  & 92.7 & 88.0 & 83.0 & 74.3 \\
  & ShieldFL   & 95.0 & 94.5 & \sbest{93.8} & 91.5  & \sbest{94.0} & \sbest{93.0} & \sbest{91.5} & 88.0  &
  \sbest{95.0} & \sbest{94.0} & \sbest{93.0} & \sbest{91.4}  & \sbest{94.5} & \sbest{94.0} & \sbest{93.0} & \sbest{91.5} \\
  & DnC        & \sbest{95.2} & \sbest{94.6} & \sbest{93.8} & \sbest{92.0}  & 93.0 & 91.0 & 89.5 & 87.5  &
  94.0 & 93.4 & 92.5 & 90.1  & 93.5 & 92.8 & 91.7 & 89.7 \\
  & DP2Guard   & \best{95.5} & \best{95.2} & \best{94.8} & \best{93.5}  & \best{95.0} & \best{93.8} & \best{93.5} & \best{92.5}  &
  \best{95.2} & \best{94.5} & \best{94.0} & \best{92.8}  & \best{95.2} & \best{94.8} & \best{94.2} & \best{93.2} \\
  \midrule

  \multirow{6}{*}{\textbf{Fashion-MNIST}}
  & FedAvg     & 73.0 & 71.4 & 68.0 & 63.3  & 64.0 & 53.4 & 32.0 & 10.0  &
  71.0 & 67.4 & 60.0 & 50.1  & 71.0 & 65.4 & 58.0 & 50.1 \\
  & FLTrust    & 76.2 & 75.8 & 75.0 & 74.2  & 75.2 & 74.5 & 73.0 & \sbest{69.7}  &
  75.2 & 74.5 & 73.5 & 69.7  & 75.0 & 74.7 & 73.5 & 70.7 \\
  & MultiKrum  & 74.7 & 74.0 & 72.2 & 68.0  & 70.7 & 63.4 & 50.2 & 34.2  &
  73.7 & 70.4 & 65.2 & 56.6  & 73.7 & 69.4 & 63.2 & 56.6 \\
  & ShieldFL   & 76.3 & 75.6 & 75.0 & 74.3  & \sbest{75.3} & \sbest{74.6} & \sbest{73.4} & 69.2  &
  \sbest{76.0} & \sbest{75.0} & \sbest{74.0} & \sbest{72.4}  & \sbest{76.3} & \sbest{75.6} & \sbest{74.4} & \sbest{72.4} \\
  & DnC        & \sbest{76.5} & \sbest{76.0} & \sbest{75.2} & \sbest{74.4}  & 74.2 & 72.2 & 70.7 & 67.9  &
  75.2 & 74.0 & 72.2 & 69.1  & 74.2 & 73.0 & 71.0 & 69.4 \\
  & DP2Guard   & \best{77.0} & \best{76.5} & \best{76.2} & \best{75.4}  & \best{76.9} & \best{76.0} & \best{75.4} & \best{74.2}  &
  \best{76.9} & \best{76.5} & \best{75.9} & \best{74.6}  & \best{76.0} & \best{75.9} & \best{75.4} & \best{74.8} \\
  \midrule

  \multirow{6}{*}{\textbf{CIFAR-10}}
  & FedAvg     & 57.0 & 54.4 & 49.0 & 40.3  & 48.0 & 35.4 & 18.0 & 10.0  &
  56.0 & 53.4 & 46.0 & 34.1  & 56.0 & 52.4 & 46.0 & 37.1 \\
  & FLTrust    & 66.5 & 65.8 & 65.0 & 63.2  & 66.0 & 64.5 & 63.0 & \sbest{60.7}  &
  66.2 & 64.5 & 63.5 & 61.2  & 66.0 & 65.2 & 64.5 & 62.2 \\
  & MultiKrum  & 63.0 & 62.0 & 59.2 & 55.0  & 58.1 & 50.8 & 38.2 & 24.3  &
  62.2 & 58.0 & 52.2 & 43.6  & 62.7 & 58.0 & 52.2 & 43.6 \\
  & ShieldFL   & 67.3 & 66.6 & \sbest{66.0} & 64.5  & \sbest{67.0} & \sbest{65.0} & \sbest{63.1} & 59.2  &
  \sbest{67.0} & \sbest{66.0} & \sbest{64.8} & \sbest{63.4}  & \sbest{67.3} & \sbest{66.2} & \sbest{65.5} & \sbest{64.0} \\
  & DnC        & \sbest{67.5} & \sbest{66.9} & \sbest{66.0} & \sbest{65.0}  & 64.1 & 62.6 & 61.0 & 58.5  &
  66.0 & 65.0 & 64.2 & 62.1  & 66.2 & 65.0 & 64.0 & 62.4 \\
  & DP2Guard   & \best{68.5} & \best{68.0} & \best{67.5} & \best{66.4}  & \best{67.4} & \best{66.5} & \best{66.0} & \best{64.5}  &
  \best{68.0} & \best{67.5} & \best{67.0} & \best{65.7}  & \best{68.0} & \best{67.5} & \best{67.0} & \best{66.2} \\

\bottomrule
\end{tabular}
\end{table*}

\subsection{Experimental Results}
\label{sec:Comparison_Analysis}

\subsubsection{Accuracy evaluation}
\label{sec:performance}
To evaluate the convergence efficiency of different aggregation algorithms, we conducted experiments over 300 global communication rounds on the MNIST, Fashion-MNIST, and CIFAR-10 datasets under a no-attack scenario, using both IID and non-IID data settings. As shown in Fig. \ref{fig:performance}, all methods achieve high final accuracy under the IID setting on MNIST and Fashion-MNIST. However, DP2Guard stabilizes within the first 50 rounds, whereas FedAvg and MultiKrum typically require over 100 rounds to converge. 
Under the non-IID setting, the performance gap widens. In particular, on the Fashion-MNIST dataset, FedAvg and MultiKrum show noticeable fluctuations during training, indicating sensitivity to data heterogeneity. 
In contrast, ShieldFL and DP2Guard exhibit smoother convergence curves. 
On the CIFAR-10 dataset with ResNet-18, all methods exhibit slower convergence due to the increased model complexity and data diversity. 
Nevertheless, DP2Guard consistently achieves competitive final accuracy and demonstrates more stable convergence compared to other baselines under both IID and non-IID settings. 

\subsubsection{Robustness}
\label{sec:robustness_evaluation} 
To comprehensively evaluate the robustness of DP2Guard, we compare all defense methods against four representative poisoning attacks: Label Flipping (LF), Fang, Min-Max, and Min-Sum, with malicious-client ratios ranging from 10\% to 40\%. 
Experiments are conducted on MNIST, Fashion-MNIST, and CIFAR-10 datasets under both IID and Non-IID settings. The results are summarized in Table \ref{tab:def_per_iid} and Table \ref{tab:def_per_noniid}.

\textbf{Performance under IID setting.} As shown in Table \ref{tab:def_per_iid}, DP2Guard consistently achieves the highest accuracy across all datasets, attack types, and malicious-client ratios. On the MNIST dataset, DP2Guard maintains accuracy above 95\% even under 40\% adversary ratio across all four attacks, while FedAvg drops sharply under Fang attack (from 90.0\% to 10.1\%) and MultiKrum degrades significantly under Min-Max and Min-Sum attacks. On Fashion-MNIST, the performance gap becomes more evident. Under the Fang attack with 40\% adversaries, FedAvg collapses to 10.0\% and MultiKrum falls to 47.2\%, whereas DP2Guard retains 78.2\%, outperforming ShieldFL (74.2\%) and DnC (73.9\%). On the more complex CIFAR-10 dataset, all methods exhibit lower overall accuracy due to the increased task difficulty. Nevertheless, DP2Guard consistently leads, achieving 72.4\% under LF attack, 71.0\% under Fang attack, 72.2\% under Min-Max attack, and 72.7\% under Min-Sum attack at 40\% adversary ratio, surpassing all baselines by 1\%--4\%. Notably, DP2Guard demonstrates the smallest accuracy degradation as the poisoning ratio increases, indicating strong resilience against escalating adversarial threats.

\textbf{Performance under Non-IID setting.} As shown in Table \ref{tab:def_per_noniid}, the Non-IID data distribution introduces additional challenges, with all methods experiencing accuracy reduction compared to the IID setting. Despite this, DP2Guard consistently maintains the best performance. On MNIST, DP2Guard achieves 93.5\% under LF attack, 92.5\% under Fang attack, 92.8\% under Min-Max attack, and 93.2\% under Min-Sum attack at 40\% adversary ratio, outperforming all baselines. The Fang attack proves most destructive in the Non-IID setting: FedAvg and MultiKrum nearly collapse on all three datasets, with accuracy dropping below 10\% and 24\% respectively on CIFAR-10 at 40\% adversary ratio. In contrast, DP2Guard retains 64.5\% accuracy under the same condition. On Fashion-MNIST and CIFAR-10, DP2Guard maintains a consistent advantage of 1\%--5\% over ShieldFL and DnC across all attack scenarios, demonstrating its effectiveness under data heterogeneity.

Overall, across all three datasets, four attack types, and both IID and Non-IID settings, DP2Guard consistently achieves the highest accuracy with the smallest performance degradation as the adversary ratio increases. These results validate the effectiveness of the hybrid detection mechanism and trust-based adaptive aggregation in DP2Guard, confirming its robustness against a wide range of poisoning threats in federated learning.

\subsubsection{Privacy}
\label{sec:privacy_risk_evaluation}
To evaluate the privacy-preserving effectiveness of DP2Guard, we conduct gradient inversion attacks using the DLG (Deep Leakage from Gradients) \cite{zhu2019deep} algorithm, which attempts to reconstruct private training data from observed gradients. Experiments are performed on both the MNIST and CIFAR-10 datasets.

\begin{figure}[t]
    \centering
    \begin{subfigure}[t]{0.48\columnwidth}
        \centering
        \includegraphics[width=\linewidth]{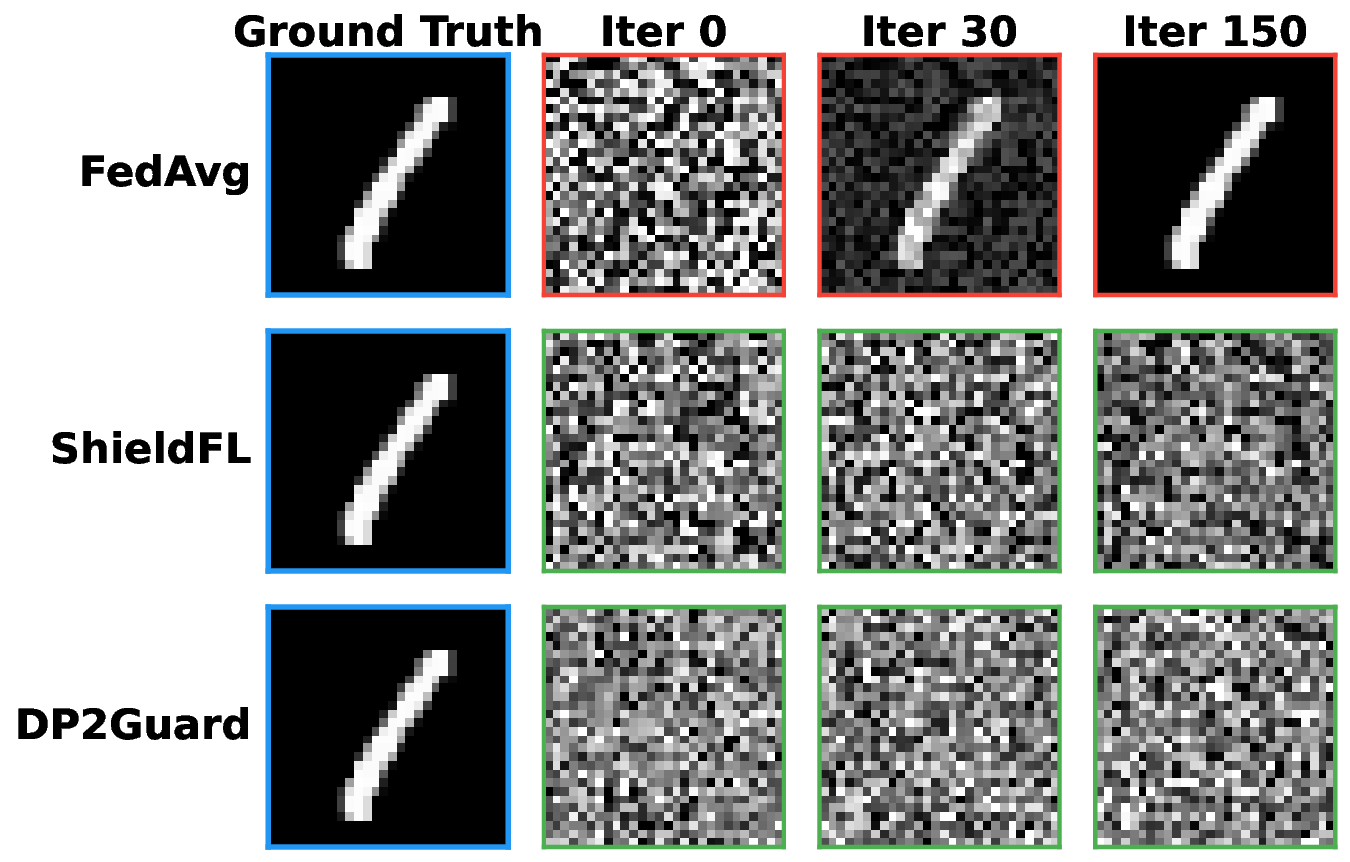}
        \caption{MNIST dataset.}
        \label{fig:gradinv_noenc}
    \end{subfigure}
    \hfill
    \begin{subfigure}[t]{0.48\columnwidth}
        \centering
        \includegraphics[width=\linewidth]{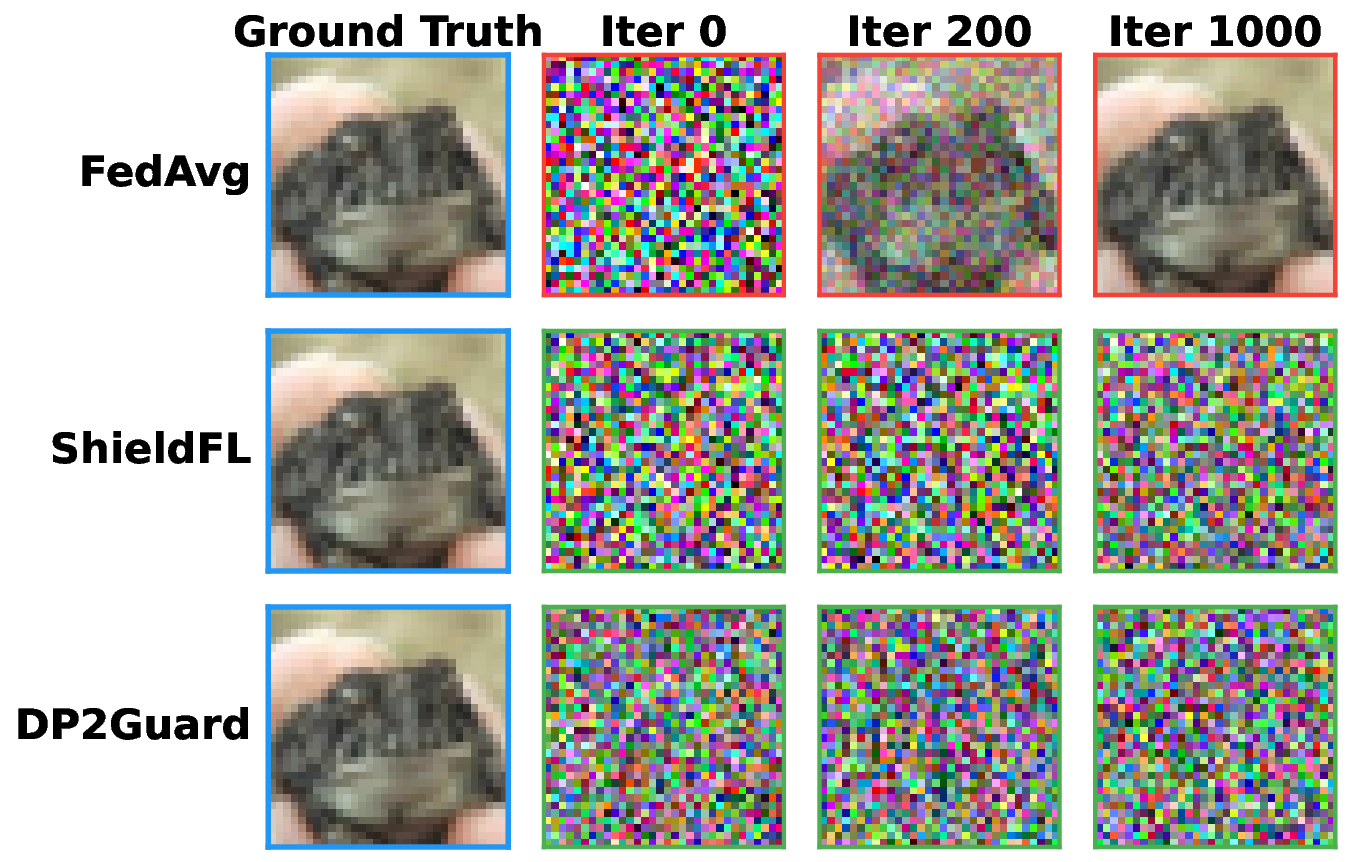}
        \caption{CIFAR-10 dataset.}
        \label{fig:gradinv_paillier}
    \end{subfigure}
    \caption{Visual comparison of images reconstructed by the DLG attack under different defense mechanisms on MNIST and CIFAR-10 datasets.}
    \label{fig:gradient_inversion}
\end{figure}

We first compare DP2Guard with ShieldFL and FedAvg under the standard honest-but-curious server model. As shown in Fig. \ref{fig:gradient_inversion}, FedAvg, which provides no privacy protection, allows the attacker to successfully reconstruct recognizable training images. 
In contrast, both ShieldFL and DP2Guard effectively prevent gradient reconstruction, with the reconstructed images being indistinguishable from random noise. These results demonstrate that the lightweight gradient splitting and masking mechanism in DP2Guard provides privacy protection comparable to the HE-based approach adopted by ShieldFL.

\begin{figure}[t]
    \centering
    \begin{subfigure}[t]{0.48\columnwidth}
        \centering
        \includegraphics[width=\linewidth]{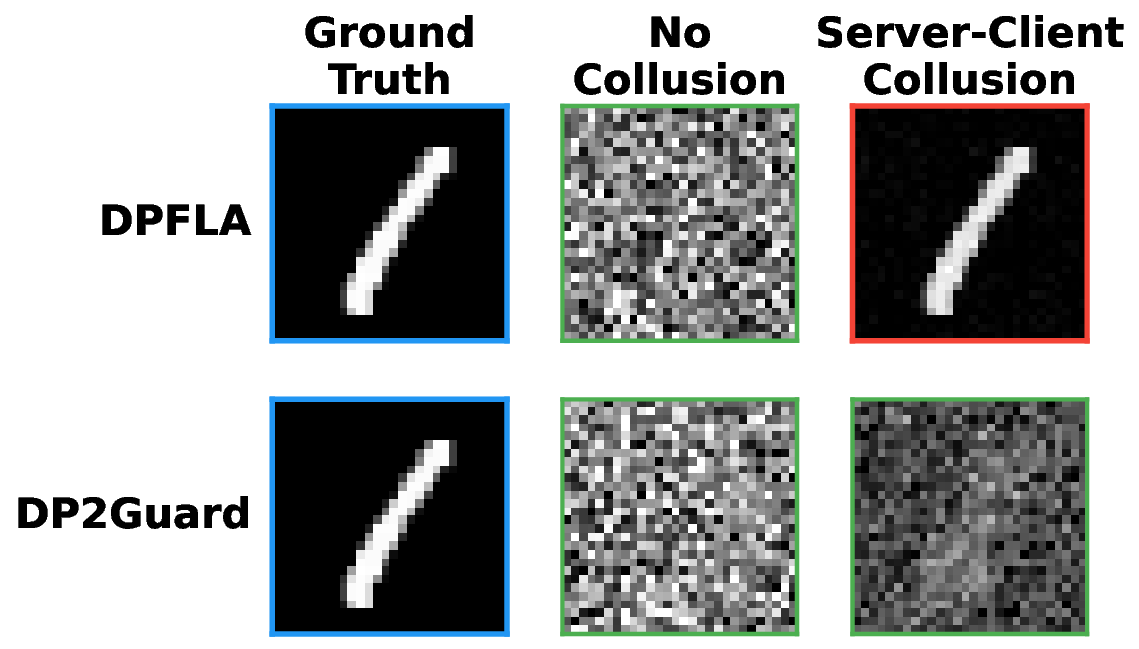}
        \caption{MNIST dataset.}
        \label{fig:gradinv_noenc}
    \end{subfigure}
    \hfill
    \begin{subfigure}[t]{0.48\columnwidth}
        \centering
        \includegraphics[width=\linewidth]{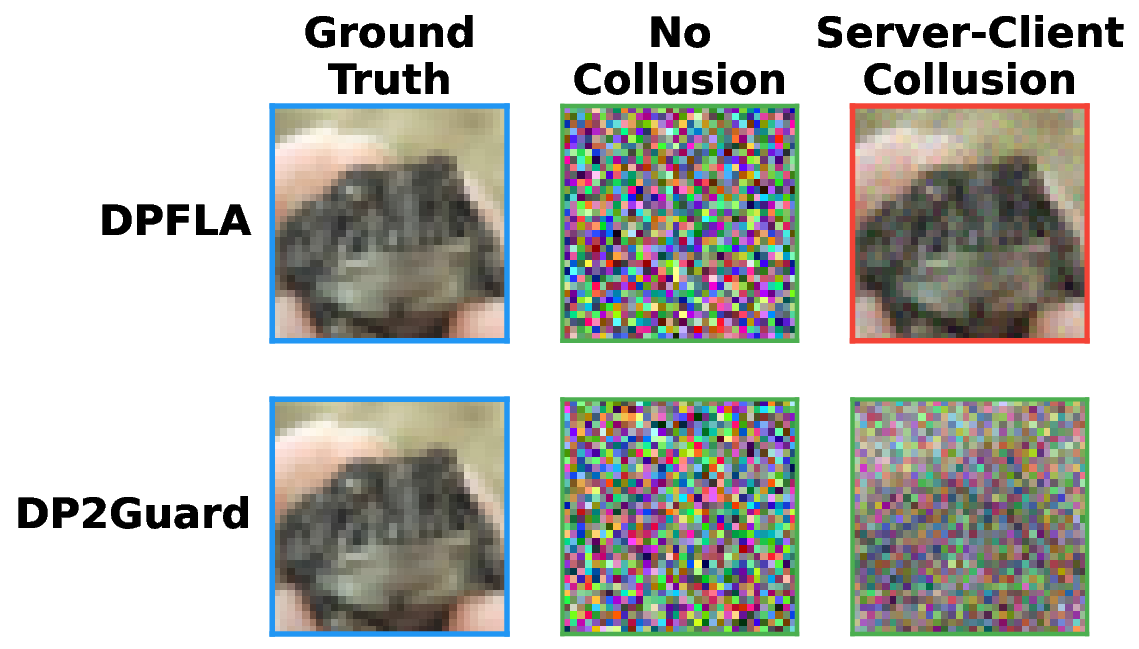}
        \caption{CIFAR-10 dataset.}
        \label{fig:gradinv_paillier}
    \end{subfigure}
    \caption{Visual comparison of images reconstructed by the DLG attack under server-client collusion between DPFLA and DP2Guard on MNIST and CIFAR-10 datasets.}
    \label{fig:gradient_inversion_collusion}
\end{figure}

We further evaluate the privacy resilience under a more challenging server-client collusion scenario, comparing DP2Guard with DPFLA. As shown in Fig. \ref{fig:gradient_inversion_collusion}, when a server colludes with a client, DPFLA fails to protect gradient privacy, and the attacker can successfully reconstruct recognizable training images. 
Under the same collusion setting, DP2Guard exhibits some residual visual features in the reconstructed images; however, the reconstruction quality is severely degraded and insufficient for a successful attack. 
This is because DP2Guard only requires non-collusion between the two servers $S_1$ and $S_2$, and does not rely on the assumption of non-collusion between servers and clients. 
These results confirm that DP2Guard provides stronger privacy protection than DPFLA even under server-client collusion.

% \begin{figure}
%     \centering
%     \includegraphics[width=0.5\linewidth]{}
%     \caption{Visual comparison for.}
%     \label{fig:gradient_inversion}
% \end{figure}

\subsubsection {Efficiency}
\label{sec:complexity_analysis}
We evaluate the efficiency of DP2Guard from both theoretical complexity and runtime overhead perspectives.

\textbf{Complexity Analysis:} Table \ref{tab:cost_analysis} summarizes the computational and communication complexity of DP2Guard compared with state-of-the-art schemes: PEFL \cite{Liu9524709}, VerifyNet \cite{XuGuowen}, and ShieldFL \cite{MaZhuoran}. 
On the client side, DP2Guard requires $\mathcal{O}\left(d T_{\text {prg}}\right)+\mathcal{O}\left(d T_{\text {add }}\right)$ computation for pseudo-random mask generation and gradient splitting, where $T_{\text {prg }}$ denotes the cost of generating a pseudo-random number and $T_{\text {add }}$ denotes the cost of a single addition operation.
These are lightweight arithmetic operations without any cryptographic overhead. 
In contrast, PEFL and ShieldFL require $\mathcal{O}\left(d T_{\text {phe}}\right)$ due to per-element PHE encryption, where $T_{\text {phe}} \gg T_{\text {add}}$, and VerifyNet incurs $\mathcal{O}\left(d N T_{\text {vss}}\right)$ due to per-element verifiable secret sharing across $N$ clients. 
For communication, DP2Guard transmits two plaintext masked gradient shares at $\mathcal{O}(2d|\omega|)$, where $|\omega|$ is the size of a single plaintext number. 
PEFL and ShieldFL transmit HE ciphertexts at $\mathcal{O}\left(d\left|\omega_{\text {phe}}\right|\right)$, where $\left|\omega_{\text {phe }}\right| \gg|\omega|$ due to ciphertext expansion. VerifyNet requires $\mathcal{O}\left((N+d N)\left|\omega_{\text {vss }}\right|\right)$ due to inter-client secret sharing. 
On the server side, the computation of DP2Guard is dominated by pairwise cosine similarity at $\mathcal{O}\left(n^2d T_{\text {add }}\right)$, performed entirely in plaintext. 
ShieldFL and PEFL perform analogous pairwise operations on HE ciphertexts at $\mathcal{O}\left(n^2 d T_{\text {phe }}\right)$, resulting in substantially higher actual computation time since $T_{\text {phe }}$ is orders of magnitude larger than $T_{\text {add}}$.

\textbf{Runtime Overhead:} To further validate the efficiency advantage of DP2Guard, we measure the actual computation time and communication cost per training round, comparing DP2Guard with PEFL, VerifyNet, and ShieldFL. 

\begin{figure}[t]
    \centering
    \begin{subfigure}[t]{0.48\columnwidth}
        \centering
        \includegraphics[width=\linewidth]{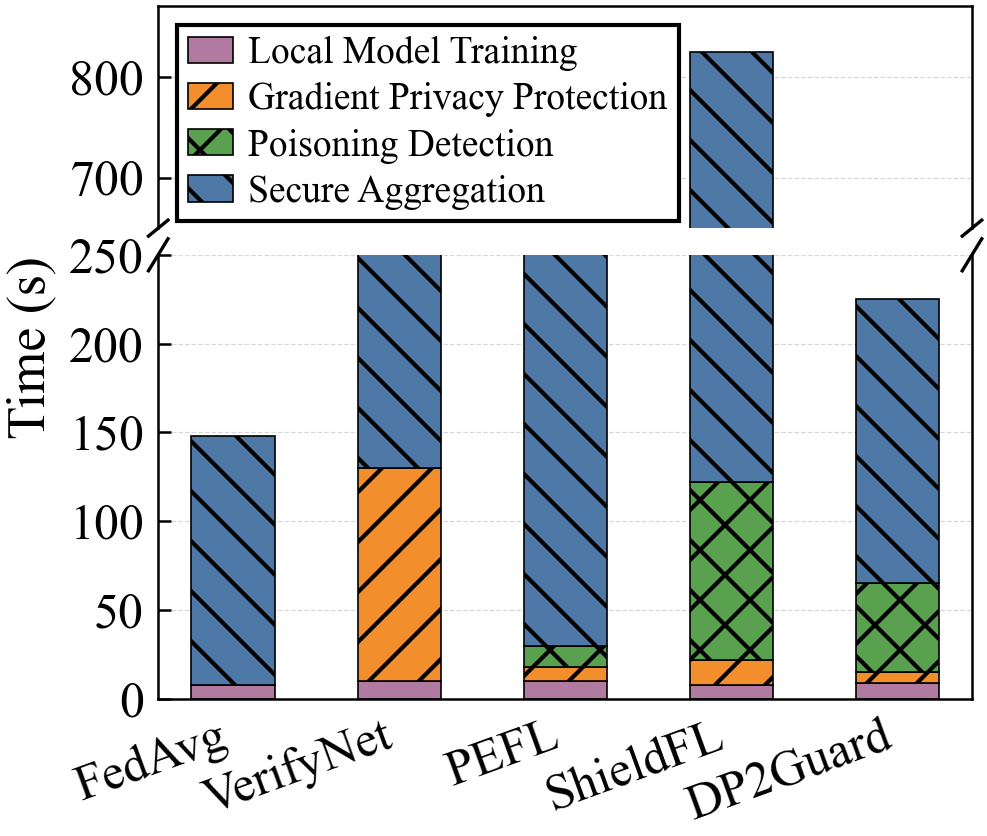}
        \caption{Computation Cost}
        \label{fig:breakdown_comp}
    \end{subfigure}
    \hfill
    \begin{subfigure}[t]{0.48\columnwidth}
        \centering
        \includegraphics[width=\linewidth]{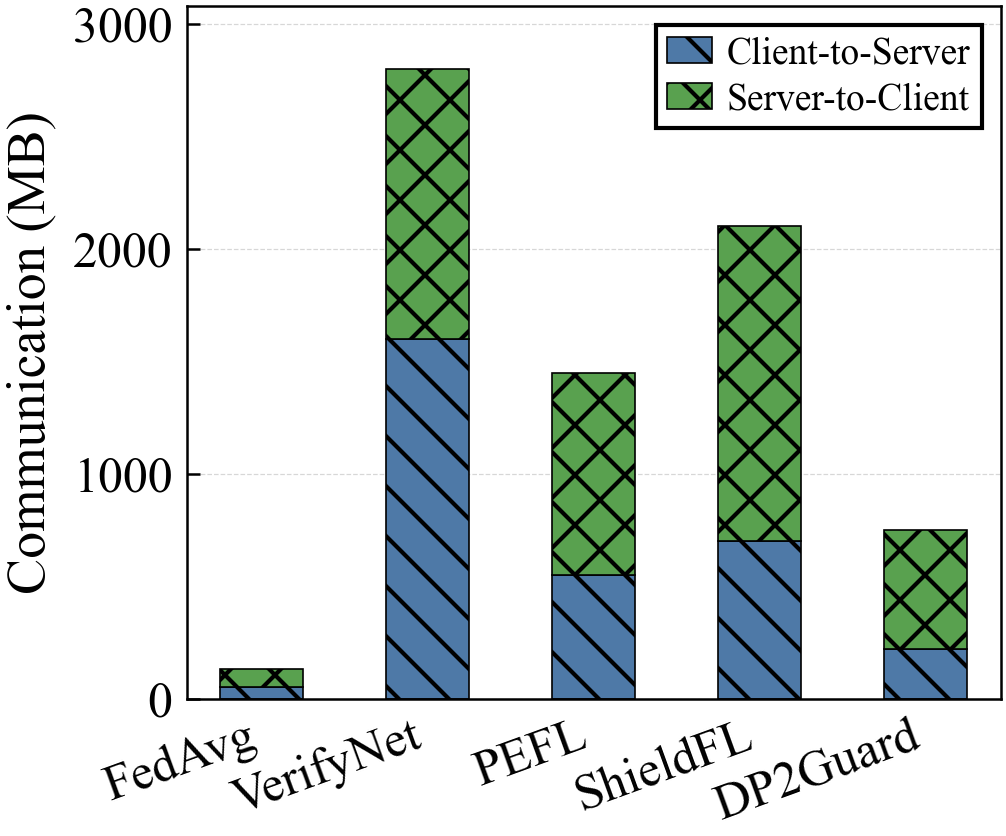}
        \caption{Communication Cost}
        \label{fig:breakdown_comm}
    \end{subfigure}
    \caption{Per-iteration overhead comparison under 40\% Byzantine clients. (a) Computational costs. (b) Communication costs.}
    \label{fig:breakdown}
\end{figure}

Fig. \ref{fig:breakdown_comp} presents the per-round execution time breakdown across different phases: local model training, gradient privacy protection, poisoning detection, and secure aggregation. The results show that ShieldFL incurs the highest overhead, followed by PEFL, both due to HE ciphertext operations in the secure aggregation phase. VerifyNet introduces significant overhead in the gradient privacy protection phase due to inter-client secret sharing and verification. In contrast, DP2Guard introduces minimal overhead beyond FedAvg, as its gradient splitting and masking operations require only lightweight plaintext arithmetic without any cryptographic operations.
Fig. \ref{fig:breakdown_comm} shows the communication overhead comparison. DP2Guard transmits plaintext masked gradient shares, resulting in substantially lower communication cost compared to VerifyNet and ShieldFL, which are dominated by inter-client secret sharing messages and HE ciphertexts, respectively. PEFL incurs moderate communication overhead due to ciphertext transmission.

\begin{table*}[t]
\centering
\small
\setlength{\tabcolsep}{3pt}
\renewcommand{\arraystretch}{1.2}
\caption{Comparison of computational and communication complexity.}
\label{tab:cost_analysis}
\begin{tabular}{lccccc}
\toprule
\textbf{Scheme} & \textbf{Technique} & \multicolumn{2}{c}{\textbf{Computation}} & \multicolumn{2}{c}{\textbf{Communication}} \\
\cmidrule(lr){3-4} \cmidrule(lr){5-6}
 & & \textbf{Server} & \textbf{Client} & \textbf{Server} & \textbf{Client} \\
\midrule
PEFL \cite{Liu9524709}
& PHE
& $\mathcal{O}(n^2 d T_{\text{phe}})$
& $\mathcal{O}(d T_{\text{phe}})$
& $\mathcal{O}(nd|\omega_{\text{phe}}|)$
& $\mathcal{O}(d|\omega_{\text{phe}}|)$ \\
VerifyNet \cite{XuGuowen}
& VSS
& $\mathcal{O}(n^2 d T_{\text{vss}})$
& $\mathcal{O}(dN T_{\text{vss}})$
& $\mathcal{O}((n^2+nd)|\omega_{\text{vss}}|)$
& $\mathcal{O}((N+d)|\omega_{\text{vss}}|)$ \\
ShieldFL \cite{MaZhuoran}
& PHE
& $\mathcal{O}(n^2 d T_{\text{phe}})$
& $\mathcal{O}(d T_{\text{phe}})$
& $\mathcal{O}(nd|\omega_{\text{phe}}|)$
& $\mathcal{O}(d|\omega_{\text{phe}}|)$ \\
DP2Guard
& GSM
& $\mathcal{O}(n^2 d T_{\text{add}})$
& $\mathcal{O}(d T_{\text{prg}}) + \mathcal{O}(d T_{\text{add}})$
& $\mathcal{O}(nd|\omega|)$
& $\mathcal{O}(2d|\omega|)$ \\
\bottomrule
\end{tabular}
\vspace{2pt}
\begin{tablenotes}
\footnotesize
\item PHE: partially homomorphic encryption; GSM: gradient splitting and masking; VSS: verifiable secret sharing. $d$: gradient dimension; $n$: number of clients. $T_{\text{prg}}$: cost of generating a pseudo-random number; $T_{\text{add}}$: cost of a plaintext addition; $T_{\text{phe}}$: cost of a PHE encryption ($T_{\text{phe}} \gg T_{\text{add}}$); $T_{\text{vss}}$: cost of generating a VSS share. $|\omega|$: size of a plaintext number; $|\omega_{\text{phe}}|$: size of a PHE ciphertext ($|\omega_{\text{phe}}| \gg |\omega|$); $|\omega_{\text{vss}}|$: size of a VSS share.
\end{tablenotes}
\end{table*}

\begin{figure}[t]
    \centering
    \begin{subfigure}[t]{0.48\columnwidth}
        \centering
        \includegraphics[width=\linewidth]{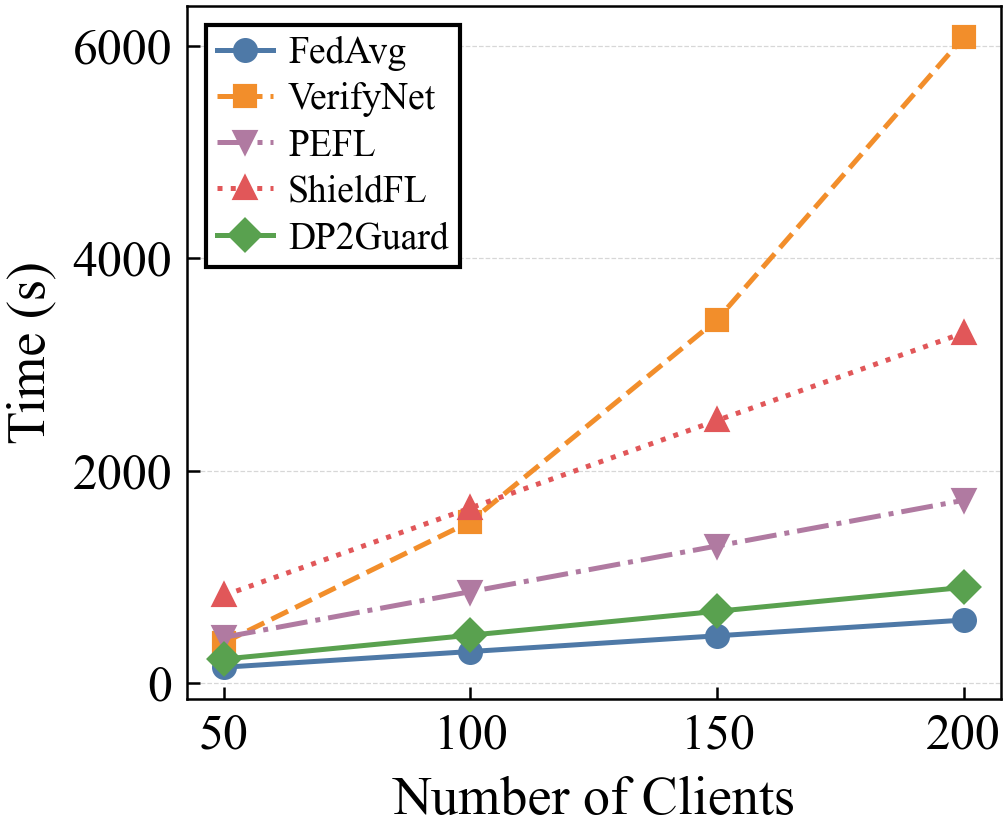}
        \caption{Computation Costs.}
        \label{fig:scalability_comp}
    \end{subfigure}
    \hfill
    \begin{subfigure}[t]{0.48\columnwidth}
        \centering
        \includegraphics[width=\linewidth]{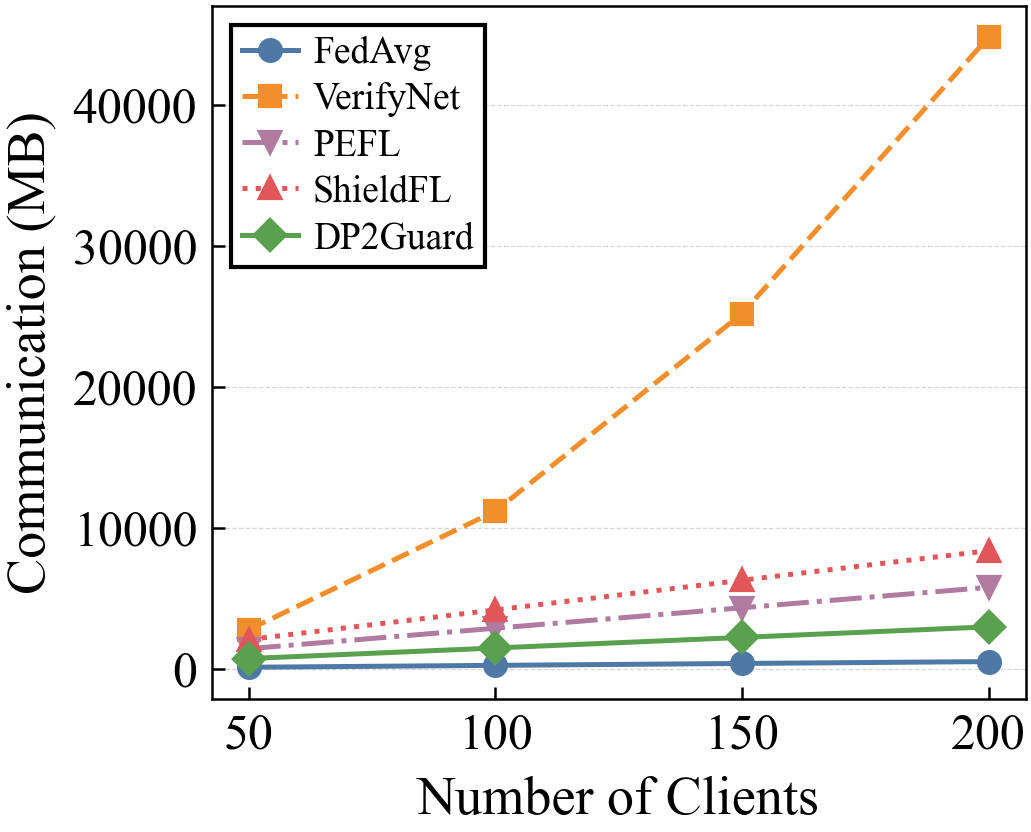}
        \caption{Communication Costs.}
        \label{fig:scalability_comm}
    \end{subfigure}
    \caption{Computational and communication overheads for different numbers of clients. (a) Computational costs. (b) Communication costs.}
    \label{fig:scalability}
\end{figure} 

\textbf{Scalability evaluation:} To further validate the practical efficiency of DP2Guard, we measure the actual computation time and communication cost under different numbers of clients $N \in\{50,100,150,200\}$. 
As shown in Fig. \ref{fig:scalability}, both the computation time and communication overhead of DP2Guard grow linearly with the number of clients, remaining close to FedAvg across all scales. In contrast, VerifyNet exhibits near-quadratic growth in both computation and communication as $N$ increases, becoming increasingly impractical at larger scales. ShieldFL and PEFL maintain moderate growth but incur substantially higher overhead than DP2Guard due to HE ciphertext operations. 
These results confirm that DP2Guard achieves a favorable balance between privacy protection and efficiency, and scales well to larger client populations.
\subsubsection{Impact of Parameter $\beta$} 
\label{sec:sensitivity_analysis}
\begin{figure}[t]
    \centering
    \begin{subfigure}[t]{0.48\columnwidth}
        \centering
        \includegraphics[width=\linewidth]{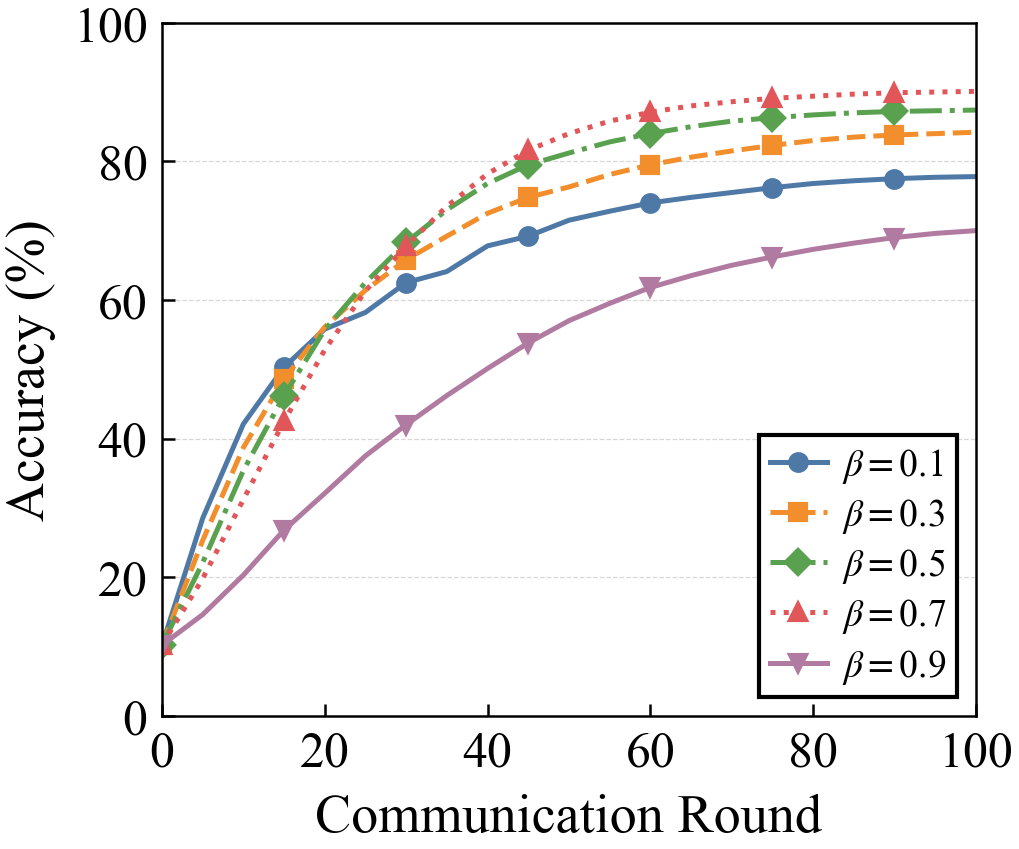}
        \caption{Convergence with different $\beta$.}
        \label{fig:beta_accuracy_rounds}
    \end{subfigure}
    \hfill
    \begin{subfigure}[t]{0.48\columnwidth}
        \centering
        \includegraphics[width=\linewidth]{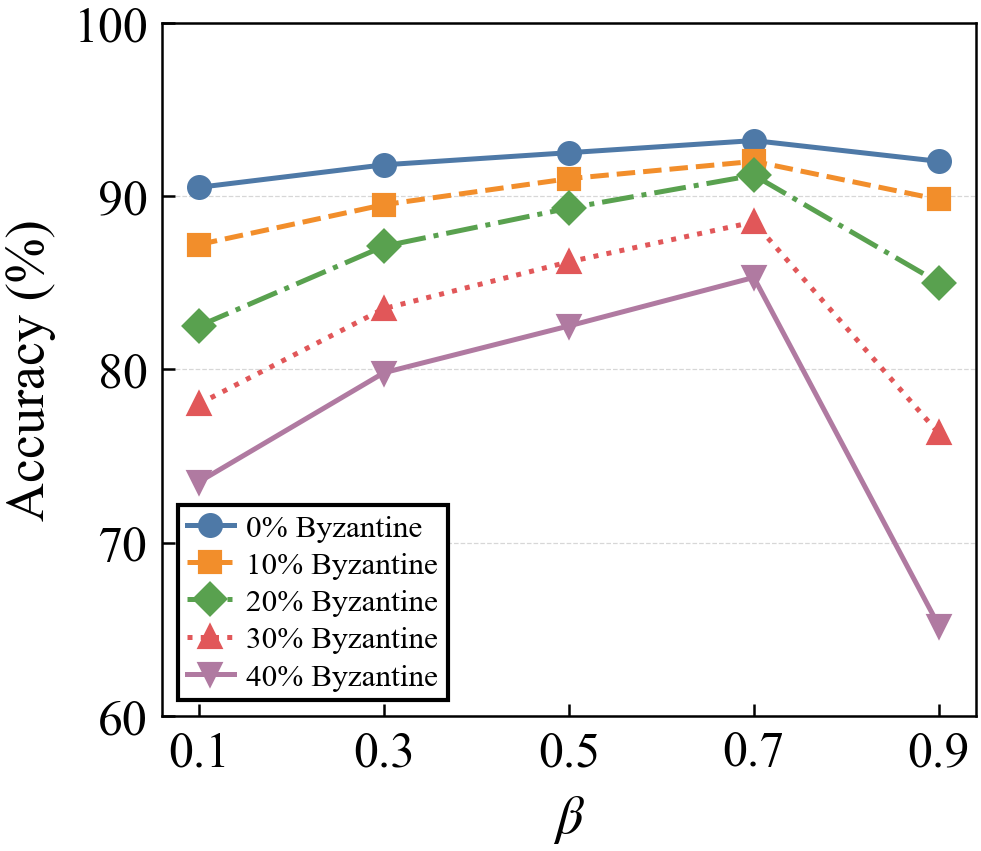}
        \caption{Effect of $\beta$ under varying Byzantine ratios.}
        \label{fig:beta_final_metrics}
    \end{subfigure}
    \caption{Impact of the factor $\beta$ on CIFAR-10. (a) Test accuracy over communication rounds  with 40\% Byzantine clients for different  $\beta$. (b) Test accuracy versus  $\beta$ under different Byzantine client ratios.}
    \label{fig:sensitivity_analysis}
\end{figure}

To investigate the sensitivity of DP2Guard to the parameter $\beta$, we conduct a parameter sensitivity analysis under different attack intensities. 
Fig. \ref{fig:beta_accuracy_rounds} illustrates the convergence curves under different $\beta$ values with a 40\% Byzantine ratio on the MNIST dataset. 
A larger $\beta$ (e.g., 0.9) assigns excessive weight to historical trust, allowing malicious clients to maintain high trust scores over extended periods, resulting in slow convergence and a final accuracy of only 70.2\%. 
A smaller $\beta$ (e.g., 0.1) responds rapidly to current-round behavior but introduces excessive trust score fluctuations, leading to unstable aggregation and a limited accuracy of 77.8\%. In contrast, $\beta = 0.7$ achieves the fastest convergence and the highest accuracy, effectively balancing responsiveness and stability. 
Fig. \ref{fig:beta_final_metrics} further demonstrates this finding across different byzantine ratios, where all curves exhibit an inverted-U shape peaking at $\beta = 0.74$. 
Based on these results, we adopt $\beta=0.7$ as the default setting across all experiments.

\section{Conclusion}
\label{sec:concludes}

In this paper, we presented DP2Guard, a lightweight PPFL framework designed to address the dual challenges of privacy leakage and vulnerability to model poisoning. To reduce computational overhead, DP2Guard replaces traditional cryptographic techniques with an efficient gradient splitting and masking strategy that protects the privacy of local model updates. To improve robustness, we designed a hybrid anomaly detection mechanism that combines cosine similarity and spectral analysis to effectively identify malicious updates, complemented by a trust scoring mechanism that penalizes both excluded and surviving anomalous clients. Furthermore, blockchain is integrated to provide a secure and auditable training process. Extensive experimental results demonstrate that DP2Guard achieves superior performance compared to state-of-the-art defenses, offering enhanced robustness and lower communication and computation costs.

% Despite these advantages, DP2Guard has certain limitations that warrant further investigation. First, the current gradient splitting and masking mechanism assumes that no global external eavesdropper can simultaneously monitor both communication channels. If such an eavesdropper exists and can bypass transport-layer encryption, it could potentially pair the two gradient shares and recover the original gradient. In future work, we plan to integrate lightweight homomorphic encryption or differential privacy techniques to ensure that gradient shares remain protected even when paired and summed. Second, under largescale partial participation settings with very low participation rates, the trust scores of inactive clients may not be updated frequently enough, reducing the effectiveness of the trust evaluation. We plan to address this by incorporating a trust score decay mechanism for inactive clients to maintain reliable trust evaluation under such scenarios.

\bibliographystyle{IEEEtran}
\bibliography{Bibliography}

\end{document}